\documentclass[axioms,article,submit,oneauthor,pdftex,12pt,a4paper]{mdpi} % for use with pdfLaTeX only
\lastpage{x}
\doinum{10.3390/------}
\pubvolume{xx}
\pubyear{2013}
\history{Received: xx / Accepted: xx / Published: xx}

\pdfoutput=1    % pdf conditional graphics
\usepackage{ifpdf}
\ifpdf
  \DeclareGraphicsExtensions{.pdf,.jpeg,.png}
\else
  \DeclareGraphicsExtensions{.eps}
\fi
\usepackage{amssymb,amsmath}
\usepackage{graphicx}
\usepackage{multirow}
\newcommand{\beq}{\begin{linenomath}\begin{equation}}
\newcommand{\eeq}{\end{equation}\end{linenomath}}
\newcommand{\bea}{\begin{linenomath}\begin{eqnarray}}
\newcommand{\eea}{\end{eqnarray}\end{linenomath}}
\newcommand{\bes}{\begin{linenomath}\begin{subequations}\begin{eqnarray}}
\newcommand{\ees}{\end{eqnarray}\end{subequations}\end{linenomath}}
\newcommand{\mbf}[1]{\mathbf{#1}}
\newcommand{\msf}[1]{\mathsf{#1}}
\newcommand{\ketbra}[2]{{\mid\! {#1} \rangle \langle {#2} \!\mid}}
\newcommand{\braket}[3]{{\langle {#1} \!\mid\! {#2} \rangle}_{#3}}

\newcommand{\abs}[1]{{\lvert {#1} \rvert}}
\newcommand{\norm}[1]{{\lVert {#1} \rVert}}
\newcommand{\mean}[2]{{\langle {#1} \rangle}_{#2}}
\newcommand{\real}{\mathrm{Re}\,}
\newcommand{\trace}{\mathrm{Tr}\,}
\newcommand{\minfty}{{-\infty}}
\newcommand{\dagr}{\dagger}
\DeclareMathOperator{\erf}{erf}

\newcommand{\revise}[1]{{#1}}
\Title{Some Notes on the Use of the Windowed Fourier Transform for Spectral Analysis of Discretely Sampled Data}

\Author{Robert W. Johnson}

\address[1]{%
Alphawave Research, 29 Stanebrook Ct., Jonesboro, GA 30238, USA}

\corres{robjohnson@alphawaveresearch.com}

\abstract{\nolinenumbers%
The properties of the Gabor and Morlet transforms are examined with respect to the Fourier analysis of discretely sampled data.  Forward and inverse transform pairs based on a fixed window with uniform sampling of the frequency axis can satisfy numerically the energy and reconstruction theorems; however, transform pairs based on a variable window or nonuniform frequency sampling in general do not.  Instead of selecting the shape of the window as some function of the central frequency, we propose constructing a single window with unit energy from an arbitrary set of windows which is applied over the entire frequency axis.  By virtue of using a fixed window with uniform frequency sampling, such a transform satisfies the energy and reconstruction theorems.  The shape of the window can be tailored to meet the requirements of the investigator in terms of time/frequency resolution.  The algorithm extends naturally to the case of nonuniform signal sampling without modification beyond identification of the Nyquist interval.}


\begin{document}\nolinenumbers%

%%%%%%%%%%%%%%%%%%%%%%%%%%%%%%%%%%%%%%%%%%%%%%%%%%%%%%%%%%%%

\section{Introduction}

The primary criticism leveled at the use of the continuous wavelet transform for the spectral analysis of discretely sampled data is that it fails to give quantitatively meaningful results.  The power spectral density produced from the convolution of a wavelet basis and a discrete signal gives a qualitative picture of the temporal variation of its frequency content; however, when one attempts the reconstruction of the signal, the residual is not on the order of the precision of one's computational device.  Likewise, the integrated margins of the power spectral density do not precisely equal the energy of the original signal.  In other words, the discrete implementation of the continuous wavelet transform and its inverse do not satisfy the energy and reconstruction theorems of spectral analysis.  The goal of this investigation is to devise a multiresolution analysis which does satisfy those theorems.

Our insistence upon the satisfaction of the energy and reconstruction theorems is because they are first principles requirements related to the conservation of physical energy.  When suitably defined, energy is not to be created nor destroyed; neither should it get lost in the shuffle.  There is a deep relation between energy content and information content, as quantum mechanics teaches us, thus a loose grip on one implies a loose grip on the other.  On a more practical level, the satisfaction of those theorems is among the requirements for a maximum entropy spectral analysis of data which includes the effects of measurement uncertainty, whose consideration is beyond the scope of this article.

The review by \citet{torrence:98} remains a popular resource for practitioners of wavelet analysis.  It relies on the method by \citet{farge-1992} for the reconstruction of the data signal.  \citet{grossman-1984} are credited with establishing the reconstruction theorem in the continuum.  \citet{Meyer-1986} and \citet{mallat-192463} developed the theory of multiresolution analysis, and \citet{daubechies-1988} constructed the first orthonormal basis with compact support, leading to the implementation of the dyadic wavelet transform in terms of finite impulse response digital filter banks~\cite{vetterli-157221}.  Some interesting applications of the continuous wavelet transform for the spectral analysis of data can be found in references~\cite{k-mmg-1987gs,meyers-1993,Baliunas:grl2411,fligge-313,ans-c49,cjaa:35391,npg-11-561-2004,piscaron-1661,liu-2093,jgr-a-2156,echer-41,greene-2273}.

This paper is organized as follows.  First, we will quickly review the theory of the continuous Fourier transform and its discrete implementation.  Next, we will look at the Gabor transform and its relation to the wavelet transform using the Morlet basis.  We will then propose an algorithm for a layered window transform whose spectral density is similar to that of the Morlet transform yet which satisfies the energy and reconstruction theorems.  Following that, we will demonstrate that the algorithm works unaltered for data with irregular sampling once the corresponding Nyquist interval is identified.  \revise{Finally, we will look at how the selection of the window affects the time/frequency resolution of the transform.}  We will conclude with a brief summary and suggestions for applications.

\revise{
Some notations used throughout this paper are explained here.  Scalars are written as $s$, while vectors and matrices are written as $\mbf{v}$ and $\msf{M}$ respectively and may be defined in terms of their components, \textit{e.g.} $\mbf{v} \equiv v_s = v(s)$.  The transpose of a vector or matrix is indicated by the superscript $\mbf{v}^T$, and the conjugate transpose by $\msf{M}^\dagr$, using the standard rules for matrix multiplication.  Inner products may be written in bra-ket notation as $\braket{a}{b}{s}$, and matrix entries as the ket-bra $\ketbra{a}{b}$.  Expectation values are notated as $\mean{v(s)}{s}$.  Sets will be indicated by their boundaries $[a, b]$, and whether continuous or discrete must be derived from context.  The operation of rounding up, \textit{i.e.} taking the next greatest integer, is denoted by $\lceil a \rceil$, and when the notation $a \rightarrow b$ is encountered, it is understood to mean ``$a$ is replaced by $b$''.  The programs used for this analysis are available from \url{http://www.alphawaveresearch.com} online.
}

%%%%%%%%%%%%%%%%%%%%%%%%%%%%%%%%%%%%%%%%%%%%%%%%%%%%%%%%%%%%

\section{Continuous and Discrete Fourier Transforms}

Let us begin by considering the Fourier transform in the continuum over axes of time and frequency.  Suppose we have some signal $y(t)$, possibly complex valued, of finite energy $E_y \equiv \braket{y}{y}{t} = \int_\minfty^\infty y^\ast(t) y(t) dt < \infty$.  If the signal carries units of $u_y$, then the signal energy has units of $u_E \equiv u_y^2 u_t$.  The units of signal energy are proportional to those of physical energy in joules by a factor of the load impedance $E_J = E_y / Z_L$; for example, if the signal has units of volts $u_y= V$ and time is measured in seconds $u_t = s$ so that $u_E = V^2 s$, then accounting for the load impedance in ohms $u_Z = \Omega$ gives physical units of $V^2 s / \Omega = (m l^2 / t^2 q)^2 t / (m l^2 / t q^2)$ in terms of mass, length, time, and charge, which is equivalent to joules.  The units of frequency are reciprocal to those of time $u_f = u_t^{-1}$, and the term ``power'' is understood to mean ``energy density'' and must be qualified by the domain for its distribution.  The squared modulus $\abs{y(t)}^2 \equiv P(t)$ can thus be identified as the temporal power of the signal.

Under suitable conditions not elaborated here, the Fourier integral and its inverse, \beq
\hat{y}(f) \equiv \int_\minfty^\infty \exp (-i 2 \pi f t) y(t) dt \; , \qquad
\hat{\hat{y}}(t) \equiv \int_\minfty^\infty \exp (i 2 \pi f t) \hat{y}(f) df \; ,
\eeq define an integral transform pair satisfying the Plancherel energy theorem $\int_\minfty^\infty \abs{y(t)}^2 dt = \int_\minfty^\infty \abs{\hat{y}(f)}^2 df$ and the Fourier inversion theorem $\hat{\hat{y}}(t) = y(t)$.  The representation over the frequency axis $\hat{y}(f)$ carries units of $u_{\hat{y}} = u_y u_t$ so that the squared modulus $\abs{\hat{y}(f)}^2 \equiv P(f)$ gives the spectral power of the signal, and the spectral energy is $E_{\hat{y}} \equiv \braket{\hat{y}}{\hat{y}}{f} = \int_\minfty^\infty \hat{y}^\ast(f) \hat{y}(f) df$.  When the signal is of finite duration $t \in [0, T]$, the Fourier transform's response to a component sinusoid $y(t) = \exp (i 2 \pi f' t)$ is \beq
\hat{y}(f) = \int_0^T \exp [-i 2 \pi (f - f') t] dt = \exp (-i \pi f_\Delta T) \dfrac{\sin (\pi f_\Delta T)}{\pi f_\Delta} \; ,
\eeq expressed in terms of the frequency offset $f_\Delta \equiv f - f'$, thus the spectral power of the uniform window function $\Phi_T(t) = T^{-1/2}$ normalized to unit energy describes the leakage in the frequency spectrum induced by the finite duration of the signal, \beq
L_T(f_\Delta) df_\Delta \equiv \abs{\hat{\Phi}_T(f_\Delta)}^2 df_\Delta = T^{-1} \left[ \dfrac{\sin (\pi f_\Delta T)}{\pi f_\Delta} \right]^2 df_\Delta \; ,
\eeq with domain $f_\Delta \in [\minfty,\infty]$.  The continuum leakage function is not periodic in $f_\Delta$, as its magnitude decays according to $f_\Delta^{-2}$.

\begin{figure}
\centering
\includegraphics[]{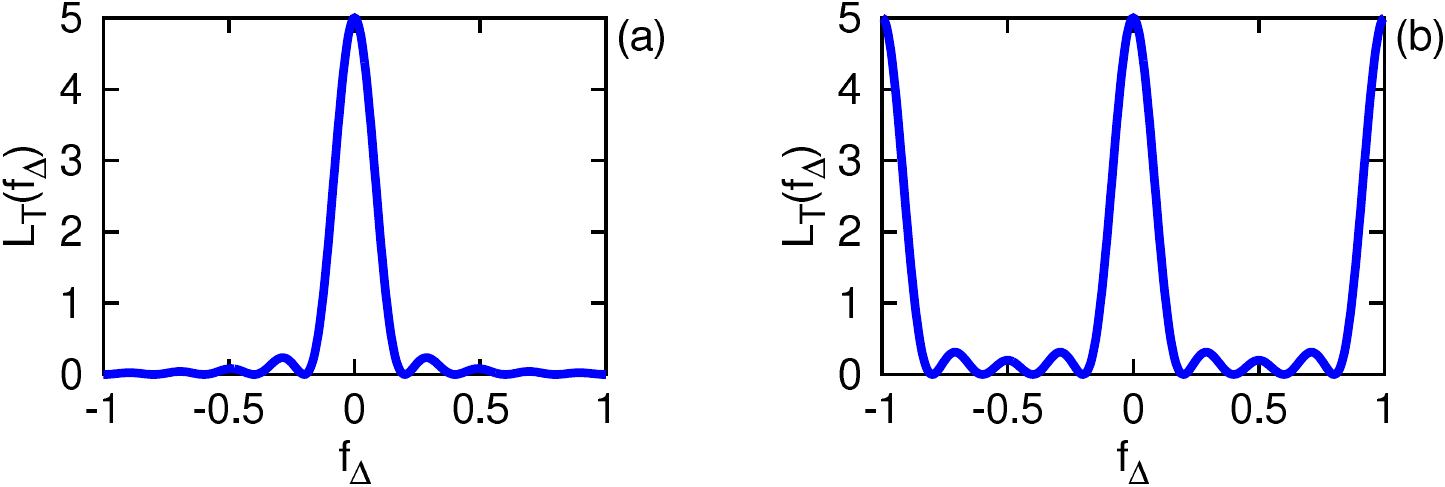}%
\caption{\label{figA} \revise{Comparison of the continuous time leakage function in (a) to its discrete time counterpart in (b) for a uniform window of duration $T = 5$ over a continuous frequency axis.}}
\end{figure}

With regard to practical data analysis, let us suppose now that the signal is given in terms of discrete samples in time with uniform duration and spacing.  (Those requirements will be relaxed in Section~\ref{sec:irreg}.)  The time axis can then be described in terms of integers $t \in [1,T]$ with unit $u_t \equiv \Delta_t$ given by the sample rate, allowing the signal to be written as a vector $\mbf{y} \equiv y_t = y(t)$.  The signal energy can be expressed in terms of matrix multiplication as $E_y = \mbf{y}^\dagr \msf{D}_t \mbf{y}$, where the temporal metric $\msf{D}_t = \msf{I}_T \Delta_t$ is proportional to the identity matrix of order $T$.  The effect of frequency aliasing induced by the uniform sampling is often misunderstood.  If we evaluate the Fourier transform of the discrete window with unit energy, \beq
\hat{\Phi}_T(f_\Delta) = T^{-1/2} \sum_{t = 1}^T \exp (-i 2 \pi f_\Delta t) \Delta_t = T^{-1/2} \exp [-i \pi f_\Delta (T+1) ] \dfrac{\sin (\pi f_\Delta T)}{\sin (\pi f_\Delta)} \Delta_t \; ,
\eeq we find that the leakage function becomes periodic in $f_\Delta$, \beq
L_T(f_\Delta) df_\Delta \rightarrow T^{-1} \left[ \dfrac{\sin (\pi f_\Delta T)}{\sin (\pi f_\Delta)} \right]^2 \Delta_t^2 df_\Delta \; ,
\eeq with period 1 in units of $u_f = \Delta_t^{-1}$.  The principle branch is usually chosen as $f \in [-1/2,1/2] \Delta_t^{-1}$, but one should always respect the periodic nature of the spectrum.  The normalization of the spectral power likewise is defined over one period of the frequency axis, $\sum_{t = 1}^T \abs{y(t)}^2 \Delta_t = \int_{-1/2}^{1/2} \abs{\hat{y} (f)}^2 df$.  In Figure~\ref{figA} we compare the continuum and discrete leakage functions for a window with duration $T = 5$, where we see the expected oscillatory lobes as well as the periodicity induced by the discrete sampling in time.

\begin{figure}
\centering
\includegraphics[]{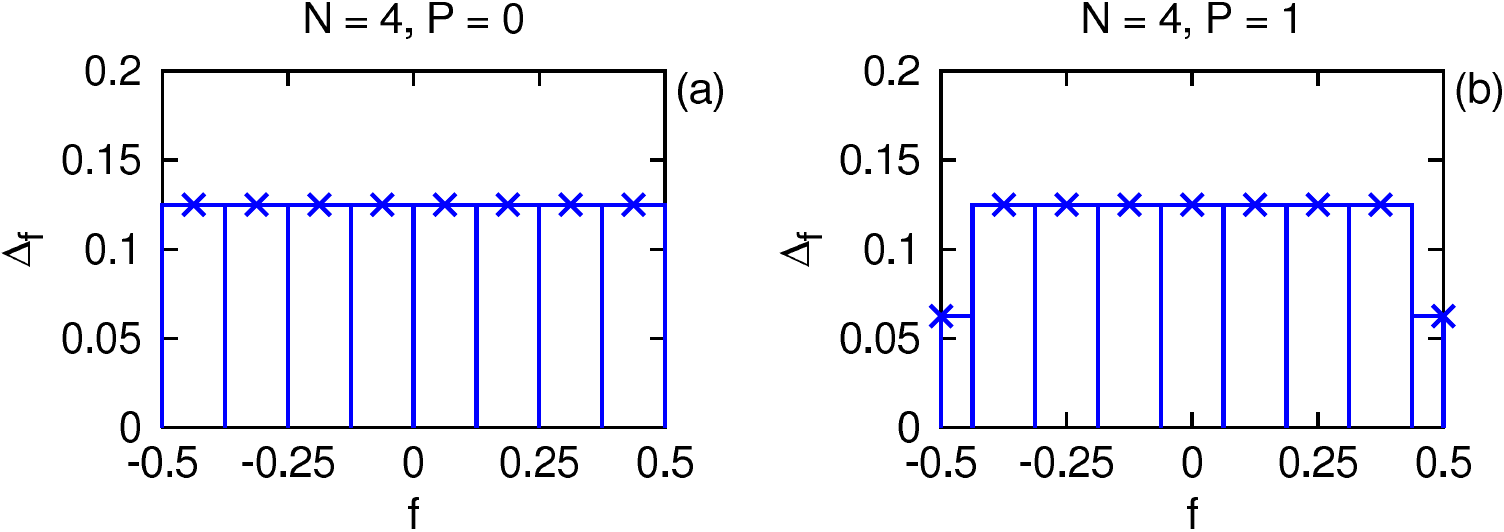}%
\caption{\label{figB} \revise{Comparison of the even (a) and odd (b) frequency discretizations of order $N = 4$ with the central frequencies indicated by $\times$.}}
\end{figure}

Now let us consider the discretization of the frequency axis into uniform bins over its principle branch $f \in [-f_c,f_c]$, where $f_c \equiv (2 \Delta_t)^{-1}$ is the Nyquist critical frequency.  To fully specify the frequency metric $\msf{D}_f$, one must state its order $N$, equal to the number of positive frequencies $\leq f_c$, as well as its parity $P \in \{0,1\}$, indicating whether an even or odd number of bins $N' = 2 N + P$ is used to span the domain.  When $P = 1$, the edgemost bins corresponding to frequencies $\pm f_c$ have a bin width equal to one half that of the others, so that the limits of integration are respected; because of aliasing, the integrand will have the same value at those locations, so that only one full bin's contribution is counted.  This convention differs from that usually given for the discrete Fourier transform~\cite{Press-1992}, where the factor of 1/2 is absorbed by the coefficients rather than the metric.  The order $N$ specifies the spacing of the frequencies as $\Delta_f = (2 N \Delta_t)^{-1}$, and the metric can be written as $\msf{D}_f / \Delta_f = \msf{I}_{N'} - (\ketbra{1}{1} + \ketbra{N'}{N'}) P / 2$ so that $\trace \msf{D}_f = \Delta_t^{-1}$.  The central frequencies of the bins can be expressed as $f_n = [n - (N' + 1)/2] \Delta_f$ for integers $n \in [1,N']$.  Figure~\ref{figB} compares the even and odd discretizations of order $N = 4$.  While the odd parity discretization is perhaps more familiar, for many of our purposes the even discretization will prove more convenient.

We are finally ready to define the two-sided discrete Fourier transform (of order $N$ and parity $P$) of a discretely sampled signal (of duration $T$).  If we collect our Fourier basis functions into the form of a matrix, $\msf{\Theta} \equiv \Theta(t,n) = \exp (i 2 \pi f_n t)$, then we can easily write the discrete transform pair in terms of matrix multiplication, $\hat{\mbf{y}} \equiv \msf{\Theta}^\dagr \msf{D}_t \mbf{y}$ and $\hat{\hat{\mbf{y}}} \equiv \msf{\Theta} \msf{D}_f \hat{\mbf{y}}$, as well as the spectral energy $E_{\hat{y}} = \hat{\mbf{y}}^\dagr \msf{D}_f \hat{\mbf{y}}$.  In component notation, one has \beq
\hat{y}(n) \equiv \sum_{t = 1}^T \exp (-i 2 \pi f_n t) y(t) \Delta_t \; , \qquad \hat{\hat{y}}(t) \equiv \sum_{n = 1}^{N'} \exp (i 2 \pi f_n t) \hat{y}(n) \Delta_f \; ,
\eeq where $\Delta_f$ is understood to account for the edge bins if $P = 1$.  If the signal $y(t)$ is real, so that $\hat{y}(-f) = \hat{y}(f)^\ast$, then one may define the one-sided transform, retaining only the nonnegative portion of the frequency axis with $N' \rightarrow N + P$ bins and $f_n \rightarrow [n - (P + 1)/2] \Delta_f$, by renormalizing the basis functions to twice the energy via $\msf{\Theta} \rightarrow \sqrt{2} \, \msf{\Theta}$ and letting $\hat{\hat{y}}(t) \rightarrow \real \hat{\hat{y}}(t)$.  Hereafter, we will assume $y(t)$ is real so that we can focus our attention on the one-sided spectrum.

In the fully discrete setting, the satisfaction of the energy and reconstruction theorems is dependent upon there being a sufficient number of degrees of freedom (dof) in the basis functions to fully represent the information content of the signal.  For real $y(t)$, the number of dof is equal to the signal duration $T$.  For the one-sided transform, whether of even or odd parity, the number of dof is equal to $2 N$, as each frequency's coefficient has both an amplitude and a phase,  $\hat{y}_n = A_n \exp (i \omega_n)$, except for the cases $f_n = 0$ or $f_c$ which have only an amplitude.  The critical frequency $f_c$ is identified as the lowest positive frequency whose basis function is entirely real over the discrete time axis, an observation which will prove useful when we consider the case of irregularly sampled data.  The minimal order beyond which the energy and reconstruction theorems are satisfied can be evaluated as $N_\mathrm{min} = \lceil T / 2 \rceil$ for a real signal (and $N_\mathrm{min} = T$ for a complex one).  That condition is realized (for even $T$) when $\msf{\Theta}^\dagr \msf{\Theta} \Delta_t = \msf{I}_{N'} T$, indicating that the basis functions $\msf{\Theta}$ form an orthogonal set for $N = T / 2$.  However, orthogonality over the discrete time axis is not a requirement, as $\mbf{y}^\dagr \msf{D}_t \mbf{y} = \hat{\mbf{y}}^\dagr \msf{D}_f \hat{\mbf{y}}$ and $\mbf{y} = \hat{\hat{\mbf{y}}}$ exactly (meaning to the precision of one's computational device) for any $N \geq N_\mathrm{min}$ and either parity $P$.

Lastly, let us look at what happens when one tries to express the fully discrete, one-sided Fourier transform over an axis of period (or scale) $s$ rather than frequency $f$.  In the continuum, the relation between the axes $f = s^{-1}$ yields the Jacobian $\abs{df/ds} = s^{-2}$, thus the spectral power over scale $P(s) \equiv \abs{\hat{y}(s)}^2$ is equal to the spectral power over frequency $P(f)$ multiplied by the Jacobian, $P(s) = f^2 P(f)$, otherwise expressed as $\hat{y}(s) = f \hat{y}(f)$.  In the discrete setting, however, one must be explicit with the mapping of the bin boundaries so that $\sum_n \abs{\hat{y}(f_n)}^2 \Delta_f = \sum_n \abs{\hat{y}(s_n)}^2 \Delta_s$.  Let $\Delta_f = b - a$ be the width of one frequency bin centered on $f_n = (a+b)/2$, which gets mapped to the scale bin $\Delta_s = a^{-1} - b^{-1}$ such that $\Delta_f / \Delta_s = a b$ and $P(s_n) = a b P(f_n)$.  One may very well ask, in what sense is $s_n = f_n^{-1}$ the center of the scale bin?  The value $f_n$ is both the mean and the median of the frequency bin with uniform measure $p^f = \Delta_f^{-1}$, but the mean period over that bin is $\mean{f^{-1}}{f} = \log (b/a) \Delta_f^{-1}$.  The value $s_n$ is recognized as the median of the scale bin with measure $p^s = s^{-2} \Delta_f^{-1}$, such that $\int_{1/a}^{2/(a+b)} p^s ds = 1/2$.  The edges for odd parity are handled similarly with respect to the half-width bins.  For either parity, the bin whose lower boundary is $a = 0$ is the one that causes trouble, as $\Delta_s = \infty$ in that case.  One can contrive to neglect that bin for odd parity by subtracting the mean of the signal $y(t) \rightarrow y(t) - \mean{y(t)}{t}$, but that procedure does not repair the problem for even parity.

\begin{figure}
\centering
\includegraphics[]{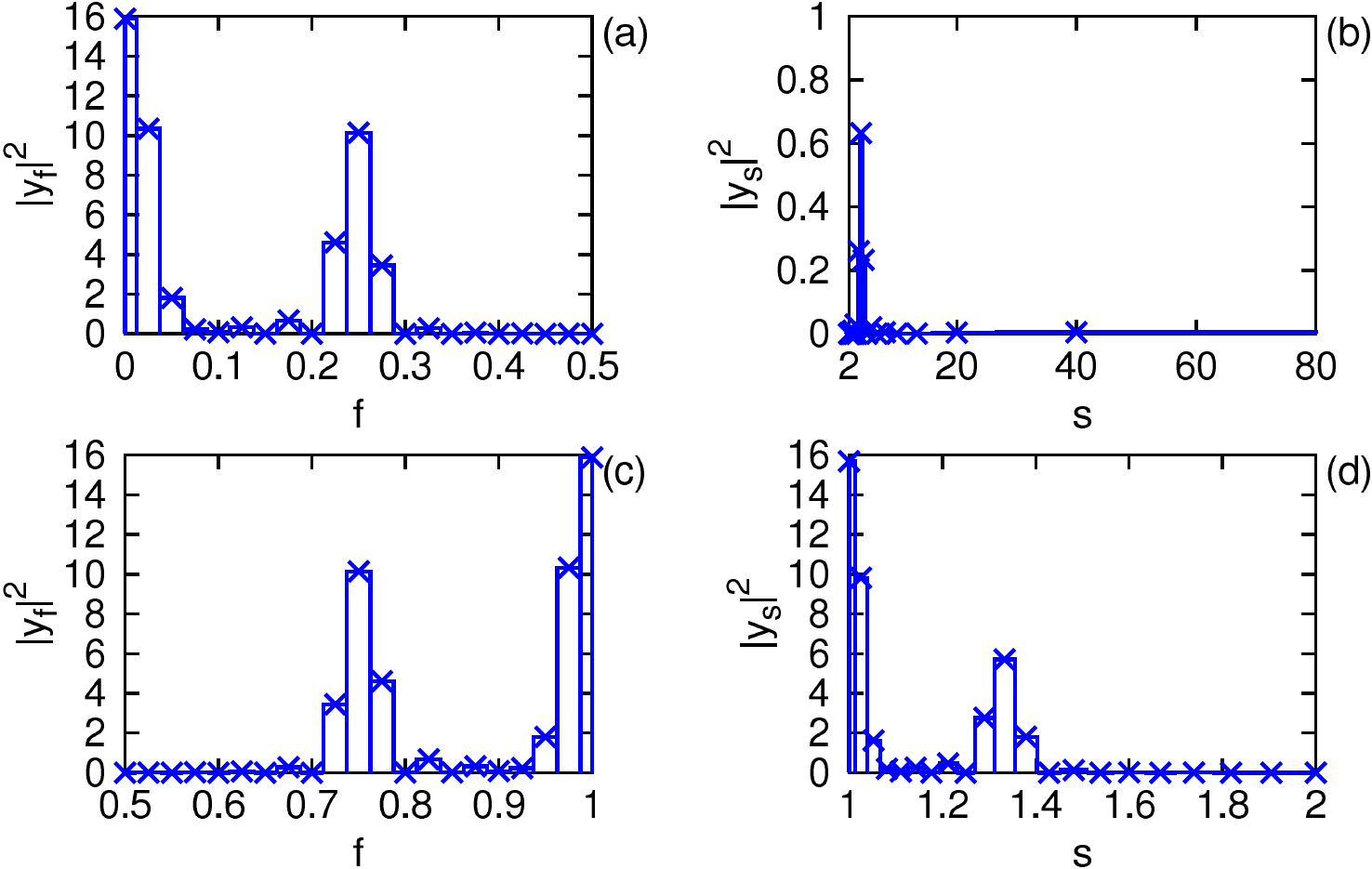}%
\caption{\label{figC} \revise{Comparison of the mapping over odd parity axes of frequency $f$ and scale $s = f^{-1}$ of the discrete spectral power of a signal with two sinusoidal components for $f \in [0,f_c]$ in (a) and (b) and for $f \in [f_c,2 f_c]$ in (c) and (d) with the central frequencies indicated by $\times$.}}
\end{figure}

To illustrate the difficulty, in Figure~\ref{figC} we display the mapping of the spectral power for a signal with duration $T = 20$, a sum of two sinusoids at frequencies 1/40 and 1/4 normalized to unit energy $E_y = 1$.  In panels (a) and (b) we show the mapping for the one-sided transform of order $N = 20$ and odd parity over domain $f \in [0,f_c]$.  While the higher frequency's contribution remains apparent, that of the lower frequency has been washed away by the measure factor.  The dof carried by the lowest frequency bin are unrecoverable from the mapping over scale on account of the infinite bin width.  In panels (c) and (d) we repeat the procedure but for domain $f \in [f_c,2 f_c]$.  This time, all the dof are properly mapped, thus the energy and reconstruction theorems are satisfied.  In Table~\ref{tabA} we give some numerical results from the evaluation shown in the figure.  In short, there is nothing to be gained by working on the scale axis rather than frequency other than a headache in dealing with the effect of aliasing, as one requires the same set of basis functions $\msf{\Theta}$ and minimal order $N_\mathrm{min}$ to satisfy the fundamental theorems of spectral analysis in either case.

\begin{table}
\centering
\begin{tabular}{c|cccc|cccc}
\hline
\multirow{2}{*}{$P$} & \multicolumn{4}{|c|}{$f \in [0,f_c]$} & \multicolumn{4}{c}{$f \in [f_c,2 f_c]$} \\
\cline{2-9}
 & $\trace{\msf{D}_f}$ & $\sum_n P(f_n) \Delta_f$ & $\trace{\msf{D}_s}$ & $\sum_n P(s_n) \Delta_s$ & $\trace{\msf{D}_f}$ & $\sum_n P(f_n) \Delta_f$ & $\trace{\msf{D}_s}$ & $\sum_n P(s_n) \Delta_s$ \\
\hline
0 & 0.5 & 1.0 & INF & NAN & 0.5 & 1.0 & 1.0 & 1.0 \\
1 & 0.5 & 1.0 & INF & NAN & 0.5 & 1.0 & 1.0 & 1.0 \\
\hline
\end{tabular}
\caption{\label{tabA} Numerical results from the evaluation of Figure~\ref{figC}.}
\end{table}

%%%%%%%%%%%%%%%%%%%%%%%%%%%%%%%%%%%%%%%%%%%%%%%%%%%%%%%%%%%%

\section{Gabor and Morlet Transforms}

Let us turn now to the consideration of how to transform the signal $\mbf{y} \equiv y(t)$ into a time-frequency representation $\hat{\msf{Y}} \equiv \hat{Y}(f,t)$ by modulating the exponential oscillations of the Fourier basis with a discrete window function, $\Theta(f,t) \rightarrow \Psi(f,t) \equiv \Phi(t) \Theta(f,t)$.  The Gabor transform is defined by the use of a Gaussian window with decay parameter $\sigma$, which we will notate as $\Phi(t) \propto \exp^{-\pi/2} (t^2/\sigma^2) \equiv e^{-\pi t^2 / 2 \sigma^2}$.  Its normalization to unit energy $\Phi \rightarrow \Phi / \braket{\Phi}{\Phi}{t}^{1/2}$ depends upon the duration of the window, parametrized by its half-width $\tau$.  In the continuum, one can write $\int_\minfty^\infty \Phi^2(t) dt \propto \sigma$ and $\int_{-\tau}^\tau \Phi^2(t) dt \propto \sigma \erf (\pi^{1/2} \tau / \sigma)$; however, in the discrete setting with index $t \in [-\tau,\tau]$, the window energy $\sum_t \Phi^2(t) \Delta_t$ is not expressible in closed form.  For the one-sided spectrum, $\Phi \rightarrow \sqrt{2} \, \Phi$ so that its energy equals 2.  To fully specify which Gabor transform is being used, one must state the values of $N$, $P$, $\sigma$, and $\tau$.  We will return to the question of finding the minimal order of the Gabor transform for a signal of duration $T$ in Section~\ref{sec:irreg}; for now, we will assume that $N$ is greater than $N_\mathrm{min}$ of the previous section so that the fundamental theorems are satisfied.

There are two alternatives for the definition of the phase convention in the transform, which differ in whether the phase is expressed relative to the origin of the signal's time axis or that of the window: \bea
\hat{Y}(n,\hat{t}) &\equiv& \sum_{t' = -\tau}^\tau \Phi(-t') \exp^{i 2 \pi} [-f_n (\hat{t} + t')] y(\hat{t} + t') \Delta_t \; , \label{eqnB} \\
\hat{\hat{y}}(t) &\equiv& \sum_{n = 1}^{N'} \exp^{i 2 \pi} (f_n t) \sum_{t' = -\tau}^\tau \Phi(t') \hat{Y}(n,t+t') \Delta_t \Delta_f \; , \label{eqnC}
\eea where the time axis for the transform coefficients carries an index $\hat{t} \in [1-\tau,T+\tau]$, or else \bea
\hat{Y}(n,\hat{t}) &\equiv& \sum_{t' = -\tau}^\tau \Phi(-t') \exp^{i 2 \pi} (-f_n t') y(\hat{t} + t') \Delta_t \; , \label{eqnD} \\
\hat{\hat{y}}(t) &\equiv& \sum_{n = 1}^{N'} \sum_{t' = -\tau}^\tau \Phi(t') \exp^{i 2 \pi} (-f_n t') \hat{Y}(n,t+t') \Delta_t \Delta_f \; . \label{eqnE}
\eea  The latter is more familiar, but either is equally valid in terms of satisfying the energy and reconstruction theorems.  The sign of the argument to the window function is chosen deliberately so that these expressions remain valid for the case of a window which is not symmetric in $t'$.  The signal is zero padded for values of $\hat{t} + t'$ outside the original domain of $[1,T]$, as no other value could be assigned without changing the energy hence information content of the signal.  For either phase convention, the temporal spectral power is defined as $P(n,\hat{t}) \equiv \abs{\hat{Y}(n,\hat{t})}^2$ with units of energy per time per frequency.  (The normalization of the window introduces a factor of $\Delta_t^{-1/2}$ so that $\hat{Y}$ carries units of $u_y u_t^{1/2}$.)  For comparison to the signal energy $E_y$, let us evaluate the spectrum's energy as $E_{\hat{Y}} \equiv \sum_n \sum_{\hat{t}} P(n,\hat{t}) \Delta_t \Delta_f$ and the reconstruction's energy as $E_{\hat{\hat{y}}} \equiv \sum_t \abs{\hat{\hat{y}}(t)}^2 \Delta_t$.

\begin{figure}
\centering
\includegraphics[]{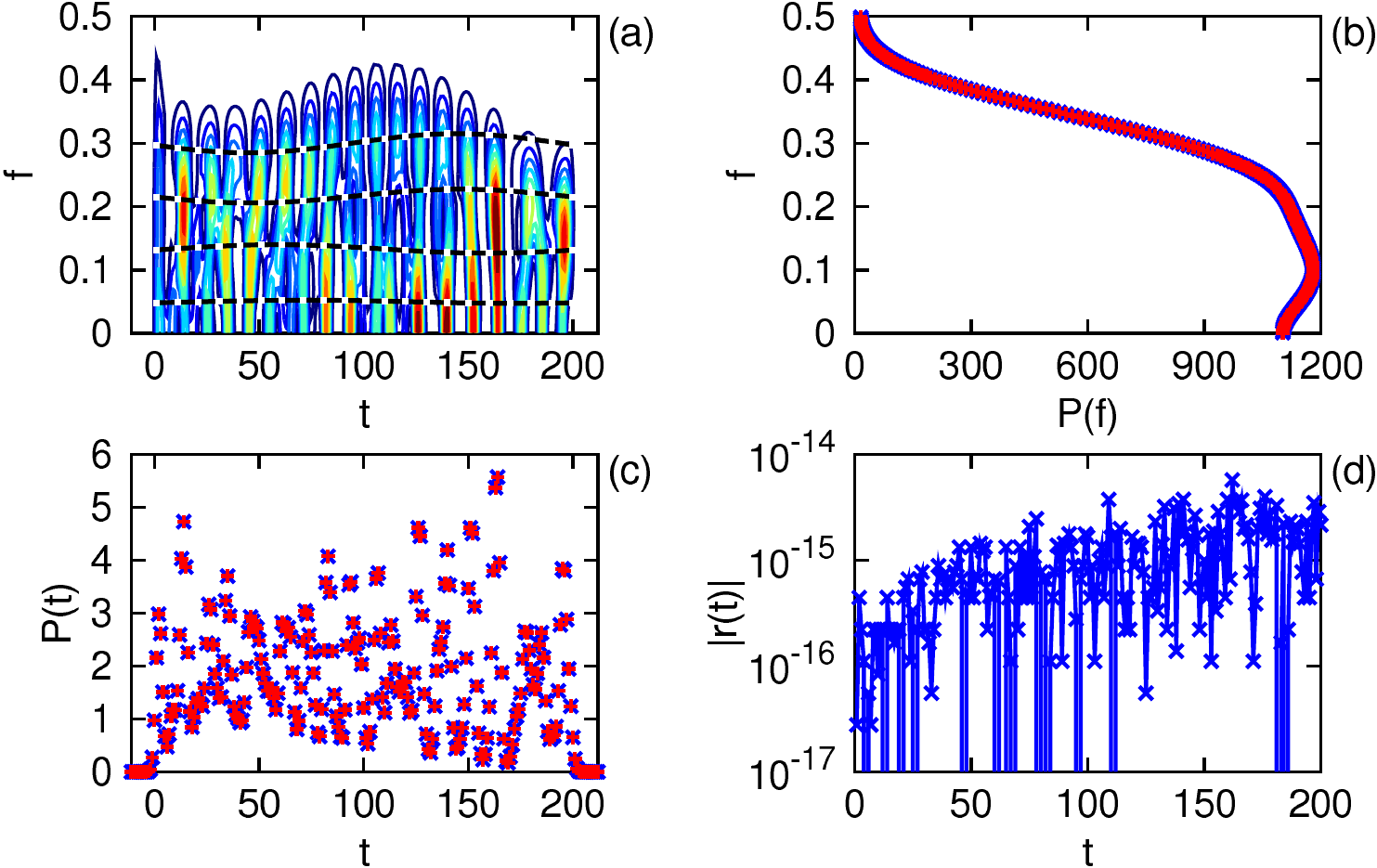}%
\caption{\label{figD} \revise{One-sided Gabor transform of a real signal as described in the text.  The power spectrum is shown in (a) with the signal's instantaneous frequencies indicated by the dashed lines.  The marginal spectral power is shown in (b) as $\times$, as is the marginal temporal power in (c), and each is compared to the convolution of the window and signal energies indicated by $+$.  The absolute value of the reconstruction residual is displayed in (d).}}
\end{figure}

For this section and the next, let us consider a particular real signal $y(t)$ comprised of 4 sinusoids with unit amplitude and non-stationary frequency of duration $T = 200$.  The variation in the instantaneous frequencies is assigned an amplitude of 5\% and a period of $T$.  Phases for the oscillations and variations are selected randomly.  In Figure~\ref{figD} we show the results of the one-sided Gabor transform with $N = 200$ and $P = 0$ using window parameters of $\sigma = 2 \sqrt{\pi}$ and $\tau = 12$.  Panel (a) displays the contours of the energy density $P(n,\hat{t})$ over axes of time and frequency.  Also indicated are the instantaneous frequencies used to generate the data.  Panels (b) and (c) display the marginal energy densities over axes of frequency and time, respectively.  The marginal energy densities are evaluated by taking each sum over an axis independently, \textit{i.e.} $P(n) \equiv \sum_{\hat{t}} P(n,\hat{t}) \Delta_t$ and $P(\hat{t}) \equiv \sum_n P(n,\hat{t}) \Delta_f$.  Panel (d) gives the absolute value of the residual $r(t) \equiv \hat{\hat{y}}(t) - y(t)$, which is on the order of the machine precision $\sim 10^{-16}$.  The ratios $E_{\hat{Y}} / E_y$ and $E_{\hat{\hat{y}}} / E_{\hat{Y}}$ are equal to unity to the same precision, thus the fundamental theorems are satisfied.  At these parameter values, the spectral resolution is poor while the temporal resolution is sharp.

The marginal densities of the Gabor power spectrum can be compared to the convolution of the window's energy density with that of the signal in either the temporal or spectral representations.  Looking first at the temporal representation, one finds the relation \beq \label{eqnF}
P(\hat{t}) = \sum_{t' = -\tau}^\tau \Phi^2(-t') y^2(\hat{t} + t') \Delta_t \; ,
\eeq where $\Phi$ in this case is normalized to unit energy.  In the spectral representation, the convolution is taken in the modular (periodic) sense, most easily performed using the two-sided transform.  Let $\hat{y}(f_n)$ be the two-sided Fourier transform of the signal with order $N$ and parity $P$, and let $\hat{\Phi}(f'_m)$ be the transform of the window function, again normalized to unit energy, with the same order and odd parity such that $M = 2 N + 1$.  Then $P(n)$ in the two-sided sense $n \in [1,2 N + P]$ can be written \beq \label{eqnA}
P(n) \approx \sum_{m = 1}^M \abs{\hat{\Phi}(M+1-m)}^2 \abs{\hat{y}[\mathrm{mod}(n + m - N - 2, 2 N) + 1]}^2 \Delta_f \; ,
\eeq where $\mathrm{mod}(a,b) \equiv a \bmod b$ and with respect to the half-width of the edge bins, from which the values corresponding to the one-sided transform can be extracted and doubled.  The relation above is written as an approximation because there is some subtlety to the evaluation of the RHS.

So far, we have imposed no condition on the window duration $T' = 2 \tau + 1$ other than being odd for integer $\tau$, and indeed, there is none.  In the lower limit $\tau = 0$, the basis functions are constant, $\Psi(0,n) = \sqrt{2}$ (or 1 for the two-sided transform), and all frequency resolution is lost, yet the fundamental theorems are still satisfied for any order $N \geq 1$.  In the opposite extreme, for example $\tau = T$ such that $T' \gg T$ with $\sigma = \infty$, the basis functions become essentially those of the Fourier transform with minimal order $N_\mathrm{min}$.  Simply put, the duration of the window in the short-time Fourier transform need not be short.  When the temporal bandwidth of the window approaches or exceeds that of the signal, the evaluation of the RHS in Equation~(\ref{eqnA}) becomes suspect if the order $N$ is not sufficient to resolve both the signal and the window; the practical solution is to increase $N$.  Furthermore, one can verify that the Gabor transform remains well behaved (if not particularly useful) for the case $\sigma \rightarrow i \sigma$ so that the window grows exponentially rather than decaying.  The only requirement is that the window be real valued and normalized explicitly, \textit{i.e.} discretely, to unit energy, times 2 for the one-sided transform.

\begin{figure}
\centering
\includegraphics[]{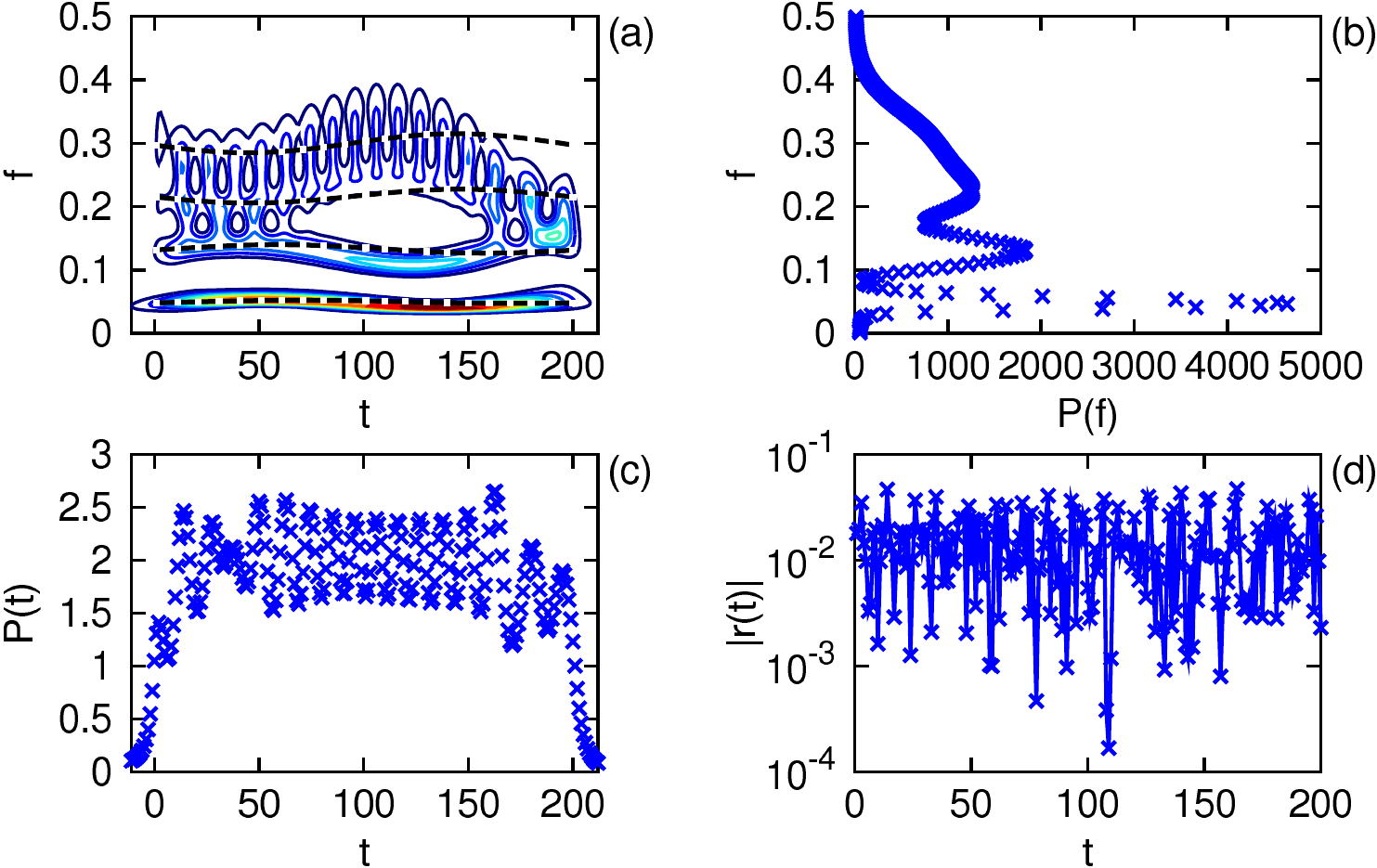}%
\caption{\label{figE} \revise{One-sided Morlet transform of a real signal as described in the text.  The power spectrum is shown in (a) with the signal's instantaneous frequencies indicated by the dashed lines.  The marginal spectral power is shown in (b) as $\times$, as is the marginal temporal power in (c).  The absolute value of the reconstruction residual is displayed in (d).}}
\end{figure}

We can now introduce the Morlet transform by promoting the window parameters from constants to functions of the central frequency, $\Phi(t) \rightarrow \Phi_n(t)$.  Let $\sigma_n \equiv \sigma / f_n$, and with respect to the discrete sampling in time, let $\tau_n \equiv \lceil \tau / f_n \rceil$ for constants $\sigma$ and $\tau$.  The parameter $\tau$ itself need not be an integer, as $\tau_n$ is what defines the window duration for bin $n$.  A difficulty with $\tau_n$ arises immediately for odd parity when $f_n = 0$; one can ameliorate the situation by using $\tau_n$ from the lowest positive bin in that case.  Again, either the phase convention of Equations~(\ref{eqnB}-\ref{eqnC}) or (\ref{eqnD}-\ref{eqnE}) can be used.  In Figure~\ref{figE} we show the results of the one-sided Morlet transform of order 200 and even parity with parameters $\sigma = \sqrt{\pi}$ and $\tau = 6$ chosen so that the window of the previous Gabor transform corresponds to the Morlet window at the critical frequency $f_c$.  The power spectrum in panel (a) has the marginal densities shown in panels (b) and (c); however, this time the ratios $E_{\hat{Y}} / E_y$ and $E_{\hat{\hat{y}}} / E_{\hat{Y}}$ differ from unity on the order of 1\%, interestingly by nearly the same value.  Likewise, the absolute value of the residual displayed in panel (d) is on the order of 1\%, comparable to the Farge method~\cite{farge-1992} as reported by Torrence and Compo~\cite{torrence:98}.  While the lowest frequency component has been well resolved, the spectral resolution of the upper frequencies remains poor.  Furthermore, without a single window in operation, one cannot perform the comparison of the marginal densities with the convolution of the window and signal energy densities corresponding to Equations~(\ref{eqnF}) and (\ref{eqnA}).

Let us insert here some comments on what is called the admissibility condition.  The statement often is made by authors that the wavelet must have a mean of zero so that its Fourier coefficient at $f = 0$ vanishes.  While that remark might apply in the continuum, it is not appropriate for discretely sampled wavelets of finite duration.  If one examines in detail the proof of the admissibility condition~\cite{Sadowsky:1996}, one finds that it relies crucially on the property of continuity.  In full, the admissibility condition states that if the wavelet is \textit{continuous} in time with \textit{continuous} Fourier transform, then its mean must be zero.  In the context considered here, neither of those conditions is met: in the temporal representation the wavelet is a piece-wise constant function with jump discontinuities at the edges of the temporal bins, and likewise for the frequency representation which, while of arbitrary resolution beyond the minimal order, must nonetheless be evaluated discretely for any practical analysis of data.  If one computes the leakage functions for the discrete Morlet basis~\cite{rwj:nova01}, one finds an artificial discontinuity is induced by the imposition of the zero mean condition.  As a final argument against subtracting the mean of the discrete basis functions, note that the Gabor transform works perfectly well for any values of the window parameters without including such procedure.

\begin{figure}
\centering
\includegraphics[]{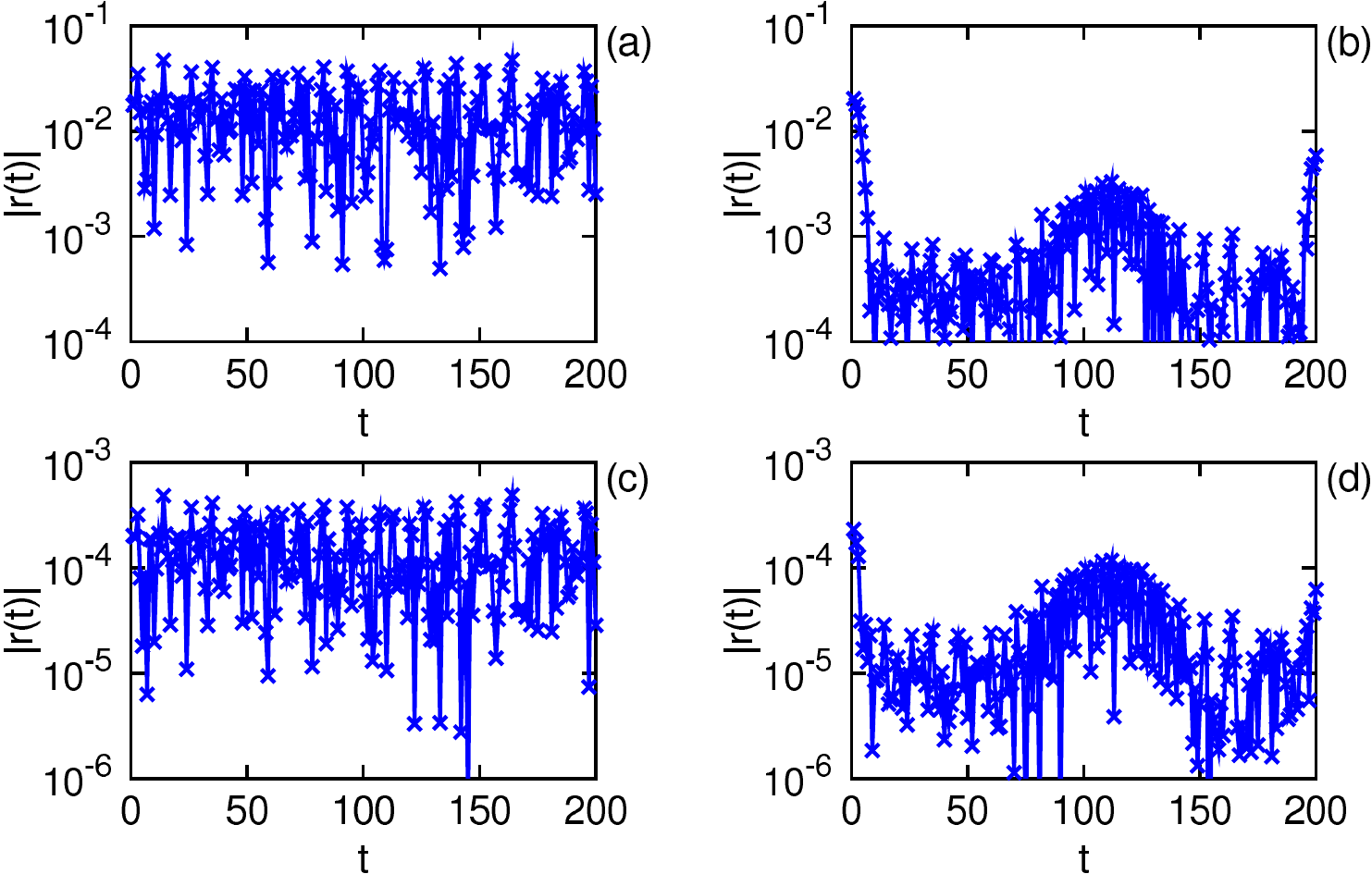}%
\caption{\label{figF} \revise{Absolute residuals for several methods of improving the transform response as described in the text.  In panel (a), the mean of the signal is subtracted; in panel (b), the mean is subtracted with renormalization of the coefficients; in panel (c), the mean is subtracted with a shift of central frequency; and in panel (d), the mean is subtracted with frequency shift and renormalization.}}
\end{figure}

There are many suggestions found in the literature for improving the performance of the discretely implemented continuous wavelet transform.  Let us examine a few of them here, with their absolute residuals displayed in Figure~\ref{figF} and a numerical summary in Table~\ref{tabB}.  For panel (a) the mean of the signal is subtracted before entering the spectral analysis.  As one can see, not much changes, but for consistency of comparison the remaining panels will also use the mean subtracted signal.  Noticing that the ratios $E_{\hat{Y}} / E_y$ and $E_{\hat{\hat{y}}} / E_{\hat{Y}}$ are very nearly equal, one can renormalize the spectrum and reconstruction \textit{a posteriori} by their geometric mean $C = (E_{\hat{\hat{y}}} / E_y)^{1/2}$, yielding the residual shown in panel (b).  This renormalization prescription works best when there is not much energy in the upper portion of the frequency axis~\cite{rwj:nova01}.  Perhaps more familiar is the prescription to shift the central frequency $2 \pi \rightarrow \omega_1$.  At these window parameter values, a shift of $\omega_1 / 2 \pi = 1.0127$ is found to work well~\cite{rwj:ijwmip01}, giving the residual shown in panel (c).  If one then applies the renormalization prescription, one gets the residual displayed in panel (d).  While there has been a noticeable improvement of three orders of magnitude, the root-mean-square residual remains far above the machine precision.

\begin{table}
\centering
\begin{tabular}{c|rrrr}
\hline
panel & \multicolumn{1}{|c}{(a)} & \multicolumn{1}{c}{(b)} & \multicolumn{1}{c}{(c)} & \multicolumn{1}{c}{(d)} \\
\hline
$\langle r^2(t) \rangle_t^{1/2}$      & 1.8215e-02 & 2.6142e-03 & 1.8213e-04 & 4.1799e-05   \\
$E_{\hat{Y}} / E_y - 1$               & 1.2885e-02 & -1.7468e-06 & 1.2674e-04 & -4.4658e-10 \\
$E_{\hat{\hat{y}}} / E_{\hat{Y}} - 1$ & 1.2889e-02 & 1.7468e-06 & 1.2674e-04 & 4.4658e-10   \\

\hline
\end{tabular}
\caption{\label{tabB} Numerical results from the evaluation of Figure~\ref{figF}.}
\end{table}

%%%%%%%%%%%%%%%%%%%%%%%%%%%%%%%%%%%%%%%%%%%%%%%%%%%%%%%%%%%%

\section{Layered Window Transform}

The primary feature of the Morlet transform is that it uses a different window function $\Phi_n$ for each bin along the frequency axis.  That is also its main difficulty, as each bin corresponds to a particular instance of the Gabor transform, any of which would work fine independently, but when the separate bins are isolated and combined, there is no reason to suppose that their marginal sums will equal the signal energy.  The correspondence between the Morlet and Gabor transforms becomes more apparent if one redefines the window decay $\sigma_n$ relative to the window width $\tau_n$ rather than the central frequency $f_n$, in which case regions along the frequency axis of the Morlet spectrum are drawn from a particular Gabor spectrum.  Simply pasting together pieces from different Gabor transforms from the outset appears doomed to failure, and it is surprising that the Morlet transform in the discrete setting works as well as it does.  While we have not given up entirely on the Morlet transform, if one wants to make quantitative use of the results of a spectral analysis, the energy and reconstruction theorems must be satisfied precisely.

How, then, are we to devise a multiresolution analysis that satisfies the fundamental theorems of spectral analysis?  The answer lies in focusing our attention on the distribution of energy in the window function.  Suppose we have one unit of energy, an ``atom'' so to speak, distributed with temporal power $\Phi_A^2(t)$ in the continuum, which we wish to use as the basis for a multiresolution analysis.  Let the window at scale $\tau$ be defined by $\Phi_\tau(t) \propto \Phi_A(t/\tau)$ for positive integer $\tau$, which scales the relative unit of time by integral amounts.  \revise{One can supplement this definition for $\tau = 0$ by letting $\Phi_0(0) = 1$ and 0 otherwise.}  For an exponential decay with scale invariant parameter $\sigma$, one can write $\Phi_\tau(t) \propto \exp^{-\pi / 2} (t^2 / \tau^2 \sigma^2)$, where the normalization to unit energy is done explicitly over the integers $t \in [-\tau,\tau]$.  (Alternately, one could hold the window width fixed for all $\tau$.)  The layered window $\Phi_L(t)$ is then defined to be the sum in terms of energy, not amplitude, of the windows $\Phi_\tau(t)$ over some set of scales $\tau \in [\tau_1, \tau_L]$ with $L$ members, normalized to unit energy, $\Phi_L^2(t) \equiv L^{-1} \sum_\tau \Phi_\tau^2(t)$.  For a one-sided transform, $\Phi_L \rightarrow \sqrt{2} \, \Phi_L$ of course.  This single window, with duration $T' = 2 \tau_L + 1$, is to be used for all the frequency bins in the windowed Fourier transform.

\revise{
Since the atomic window $\Phi_A$ is arbitrary, the distinction between summing the window energies $\Phi_\tau^2$ \textit{versus} summing the window amplitudes $\Phi_\tau$ is really just a matter of interpretation.  However one defines the sum over scales, the layered window eventually is normalized explicitly to one unit of energy, or two units for a one-sided transform.  Using the construction above in terms of energy, the amplitude of the layered window is the root-mean-square of the amplitudes of the scaled windows, while the construction summing the window amplitudes is not so simply normalized.  With respect to its physical motivation, energy is the quantity that ultimately gets quantized, not amplitude, which is a property reflected in our favored definition of the layered window.  The selection by the investigator of the set of scales $\{ \tau \}$ to use in the construction specifies the domain of energy distributions to which the transform will be sensitive.
}

\begin{figure}
\centering
\includegraphics[]{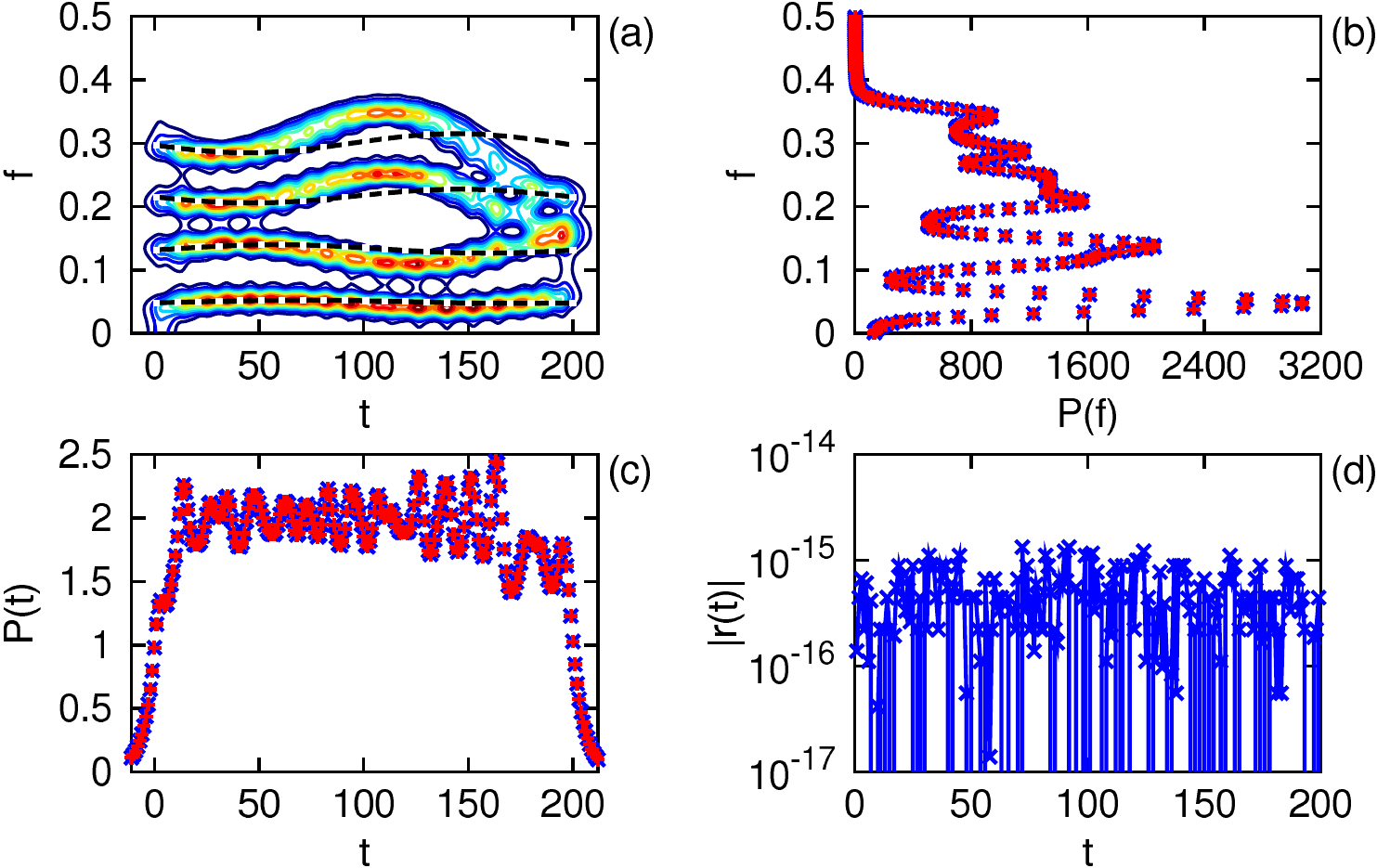}%
\caption{\label{figG} \revise{One-sided layered window transform of a real signal as described in the text.  The power spectrum is shown in (a) with the signal's instantaneous frequencies indicated by the dashed lines.  The marginal spectral power is shown in (b) as $\times$, as is the marginal temporal power in (c), and each is compared to the convolution of the window and signal energies indicated by $+$.  The absolute value of the reconstruction residual is displayed in (d).}}
\end{figure}

Let us return now to consideration of our four frequency test signal, with its mean restored.  For comparison with the previous Gabor and Morlet transforms, let the decay parameter here be $\sigma =\sqrt{\pi} / 6$, and let $\tau \in [12, 96]$.  The results of such analysis are displayed in Figure~\ref{figG}.  Compared to the Morlet transform, the spectral resolution is much improved in the upper portion of the frequency axis, as seen in panel (a).  Since only one window is in operation, the marginal densities of panels (b) and (c) can be compared to the convolution of the window and signal energy densities in either the temporal or spectral representations.  The absolute residual shown in panel (d) is on the order of the machine precision, with maybe a slight accumulation of truncation errors.  The ratios $E_{\hat{Y}} / E_y$ and $E_{\hat{\hat{y}}} / E_{\hat{Y}}$ equal unity to machine precision.  The layered window transform, by virtue of using a single window for all frequencies, satisfies the energy and reconstruction theorems to the accuracy of one's computational device.

\begin{figure}
\centering
\includegraphics[]{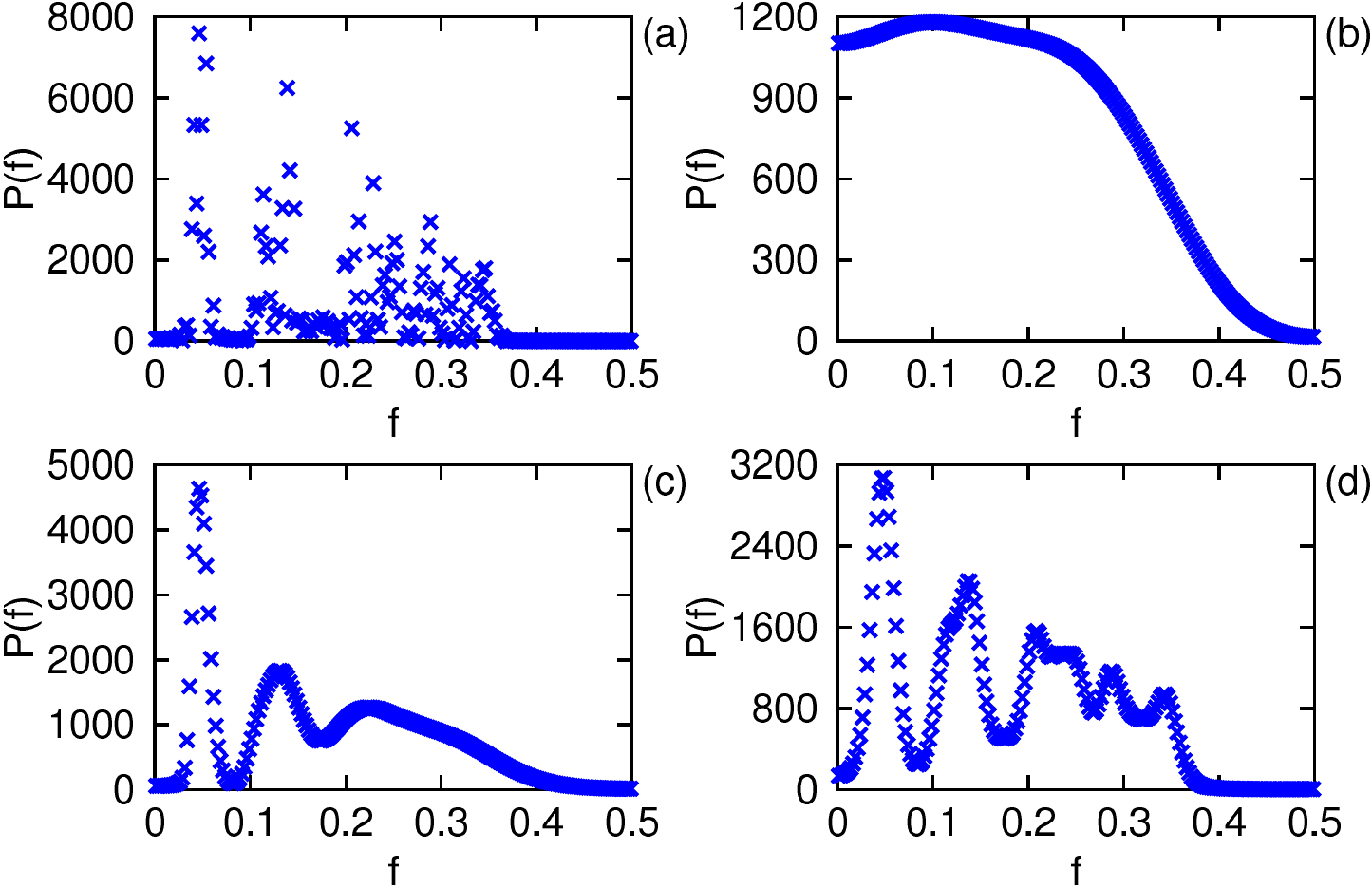}%
\caption{\label{figH} \revise{Comparison of the Fourier power spectrum in (a) to the marginal spectral energy densities of the Gabor transform in (b), the Morlet transform in (c), and the layered window transform in (d).}}
\end{figure}

As a final comparison, let us put the Fourier power spectrum of the signal alongside the marginal spectral energy densities of the Gabor, Morlet, and layered window transforms, displayed in Figure~\ref{figH}.  The Fourier spectrum in panel (a) is the typical mess one gets when analyzing non-stationary signals, but its net energy does equal the signal energy.  The Gabor spectrum in panel (b) has lost its frequency resolution on account of the tight temporal bandwidth of the selected window parameters yet retains the correct normalization.  The Morlet spectrum in panel (c) resolves the low frequency portion of the spectrum but not the high frequency region and does not have the correct normalization.  The layered window spectrum in panel (d) is normalized to the signal energy and resolves both the low and high frequency parts of the spectrum while maintaining sufficient temporal resolution to be useful in identifying non-stationary features in the signal.

%%%%%%%%%%%%%%%%%%%%%%%%%%%%%%%%%%%%%%%%%%%%%%%%%%%%%%%%%%%%

\section{Irregular Sampling and Minimal Order}
\label{sec:irreg}
 
Let us now consider the case of an irregularly sampled signal $y(t)$, with regard to the analyses by \citet{Lomb-447} and \citet{Scargle-835}.  Irregular sampling has long vexed practitioners of wavelet analysis, leading to a variety of suggestions for how to overcome its difficulties.  Because of the nature of stellar observations, the astrophysical community often presents data which contains gaps or is otherwise on an irregular time axis.  The method by \citet{Foster:1996} focuses on the normalization of the gapped wavelet basis functions, while the method by \citet{Frick:1997426} focuses on the admissibility condition; taken together, those ideas lead to the edge adapted algorithm proposed by \citet{rwj:astro01}.  More familiar perhaps is the lifting scheme by \citet{sweldens98lifting}, which operates in the context of the dyadic wavelet transform implemented in terms of finite impulse response digital filter banks.

For simplicity, we will assume that the samples each have a uniform duration $\Delta_t$ which is not greater than any of the intermeasurement periods, so that the time metric $\msf{D}_t$ remains proportional to the identity matrix; otherwise, a suitably generalized $\msf{D}_t$ must be used.  Let the observation times be given by the vector $\mbf{t} \equiv t_d$ indexed by $d \in [1,D]$, where $D$ is the total number of samples, in some unit $u_t$ not necessarily equal to $\Delta_t$ such that the values $t_d$ are integers; $u_t$ is the resolution of the time measuring apparatus, thus $\Delta_t$ must also be an integer.  Similarly, the measurements can be written as the vector $\mbf{y} \equiv y_d$ so that the signal energy remains $E_y = \mbf{y}^\dagr \msf{D}_t \mbf{y}$ in units of $u_y^2 u_t$.  Missing values, \textit{i.e.} measurements at integer times not in $\mbf{t}$, are effectively treated as zero.  Let us look first at how the discretized Fourier transform is modified in this case.

To identify the Nyquist critical frequency, one must find the lowest positive Fourier basis function which is entirely real (technically up to a constant phase) over the given set of observation times~\cite{rwj:astro03}.  That procedure is easily accomplished by shifting and scaling the time axis $\mbf{t} \rightarrow \mbf{t}'$ such that $f_c \equiv (2 u_{t'})^{-1}$ in the new time units.  If one defines $t_g$ to be the greatest common divisor over the set of intermeasurement periods, the new time axis can be written $t_d' \equiv (t_d - t_1) / t_g + 1$ in units of $u_{t'} \equiv t_g u_t$ such that $t_d' \in [1,T]$ remains integer valued.  At the critical frequency, the Fourier basis function is $\exp^{i \pi} (2 f_c t_d') = \pm 1$, and the Fourier spectrum has period 1 in units $u_f = u_{t'}^{-1}$.  Having found the units in which $f_c = 1/2$, let us drop the prime distinguishing the scaled time axis in the following, and for convenience let us suppose $\Delta_t = 1$ in the scaled units; otherwise one must be extra careful with the scaling of the energy units.

The next task is to determine the minimal order for satisfaction of the energy and reconstruction theorems.  Let $N_D \equiv \lceil D/2 \rceil$, and let $N_T \equiv \lceil T/2 \rceil$.  For $N < N_D$ there are an insufficient number of dof in the discrete Fourier transform to fully represent the information content of the (real) signal, thus the fundamental theorems are not satisfied in general.  Likewise, for $N \geq N_T$ there are more than enough dof to represent the signal; $N_T$ gives the minimal order of the analogous regularly sampled signal where the missing entries are assigned the value zero.  For $N_D \leq N < N_T$, the Fourier transform might work, depending upon the particulars of the irregular sampling.  To satisfy the reconstruction theorem $\hat{\hat{\mbf{y}}} = \mbf{y}$ for $\hat{\hat{\mbf{y}}} \equiv \msf{Q} \mbf{y}$ in the one-sided case, one must have $\msf{Q} \equiv \real \msf{\Theta} \msf{D}_f \msf{\Theta}^\dagr \msf{D}_t = \msf{I}_D$ to the order of machine precision; the energy theorem also is satisfied when that condition is met.  (For a complex signal, the entire matrix, not just its real part, must be utilized, and of course $\msf{D}_f$ must cover one full period of the frequency axis.)  The basis functions are evaluated only at the observation times $\Theta(d,n) = \exp^{i 2 \pi} (f_n t_d)$, because those are the only locations at which the reconstruction theorem must be satisfied.  In Table~\ref{tabC} we compare the norm of the reconstruction residual $\mbf{r} \equiv \hat{\hat{\mbf{y}}} - \mbf{y}$ for some signal with $D = 10$ and $T = 28$ to the distance between the quality matrix $\msf{Q}$ and the identity $\msf{I}_D$ according to the Frobenius metric at various orders $N$ with $P = 0$.  To produce a smooth picture of the spectral content, one should take $N \gg N_T$.

\begin{table}
\centering
\begin{tabular}{c|cccccccccc}
\hline
$N$ & \multicolumn{1}{c}{5} & \multicolumn{1}{c}{6} & \multicolumn{1}{c}{7} & \multicolumn{1}{c}{8} & \multicolumn{1}{c}{9} \\ \hline     
$\norm{\mbf{r}}_F$             & 3.5943e+00 & 5.1760e+00 & 3.0368e+00 & 5.5110e-15 & 3.0148e+00 \\                                        
$\norm{\msf{Q} - \msf{I}_D}_F$ & 2.4495e+00 & 2.8284e+00 & 2.4495e+00 & 5.6505e-15 & 2.0000e+00 \\ \hline                                 
$N$ & \multicolumn{1}{c}{10} & \multicolumn{1}{c}{11} & \multicolumn{1}{c}{12} & \multicolumn{1}{c}{13} & \multicolumn{1}{c}{14} \\ \hline
$\norm{\mbf{r}}_F$             & 6.6842e-01 & 5.4143e-15 & 2.6771e+00 & 6.2415e-15 & 7.4742e-15 \\                                        
$\norm{\msf{Q} - \msf{I}_D}_F$ & 1.4142e+00 & 4.3648e-15 & 2.0000e+00 & 4.2983e-15 & 4.9648e-15 \\                                        

\hline
\end{tabular}
\caption{\label{tabC} Comparison of the norm of the residual of the Fourier transform to the distance from the quality matrix $\msf{Q}$ to the identity $\msf{I}_D$ at various orders $N$ and even parity for an irregularly sampled signal with $D = 10$ and $T = 28$ \revise{such that $N_D = 5$ and $N_T = 14$.}}
\end{table}

Turning now to consideration of the Gabor transform (and by extension all windowed Fourier transforms with a fixed $\Phi$), the construction of the quality matrix is a bit more complicated, owing to the convolutions along the time axis.  It is commonly remarked that the number of dof along the frequency axis is multiplied by the number of times the window duration $T' = 2 \tau + 1$ fits within the signal duration $T$, but the situation is more subtle than that, as one must also account for the \revise{bandwidth of the window given by the decay $\sigma$.}  Beginning with the case of regular sampling $D = T$, using the phase convention of Equations~(\ref{eqnD}-\ref{eqnE}), one can write the composition of the inverse and forward transforms as \bes
\hat{\hat{y}}(t) &=& \real \sum_{n = 1}^{N'} \sum_{t' = -\tau}^\tau \Phi(t') \exp^{i 2 \pi} (-f_n t') \sum_{t'' = -\tau}^\tau \Phi(-t'') \exp^{i 2 \pi} (-f_n t'') y(t + t' + t'') \Delta_t \Delta_t \Delta_f \\
 &=& \real \sum_{t''' = -2 \tau}^{2 \tau} \sum_{t' + t'' = t'''} \Phi(t') \Phi(-t'') \sum_{n = 1}^{N'} \exp^{i 2 \pi} (-f_n t''') y(t + t''') \Delta_f \Delta_t \Delta_t \\
 &\equiv& \sum_{t''' = -2 \tau}^{2 \tau} q(t''') y(t + t''') \Delta_t \; ,
\ees which yields one row in the quality matrix $\msf{Q}$ indexed by $t''' \in [-2 \tau, 2 \tau]$ indicating the diagonal where $q(t''')$ appears such that $\msf{Q}$ has the form of a Toeplitz matrix of rank $T$.  For a symmetric window $\Phi(-t') = \Phi(t')$ written as a diagonal matrix $\msf{\Phi}$ such that $\msf{\Psi} \equiv \msf{\Phi} \msf{\Theta}$, the values $q(t''')$ can be extracted from $\msf{\Psi}^\ast \msf{D}_f \msf{\Psi}^\dagr$.  For an irregularly sampled signal $T > D$, one retains only those rows and columns of $\msf{Q}$ corresponding to actual data entries, $\msf{Q} \rightarrow \msf{Q}(\mbf{t},\mbf{t})$.  In Table~\ref{tabD} we compare the residual norm of the Gabor transform with $\tau = 12$ and two values of $\sigma$ to the distance from $\msf{Q}$ to $\msf{I}_D$ at various $N$ with $P = 0$ for an irregularly sampled signal of duration $T = 100$ with $D = 30$ entries.  \revise{As $N$ approaches $N_D$ from below, the quality of reconstruction improves until the residual is on the order of machine precision.}

\begin{table}
\centering
\begin{tabular}{c|c|cccccccccc}
\hline
$\sigma$ &                        $N$ & \multicolumn{1}{c}{6} & \multicolumn{1}{c}{7} & \multicolumn{1}{c}{8} & \multicolumn{1}{c}{9} & \multicolumn{1}{c}{10} \\ \hline         
\multirow{5}{*}{$2 \sqrt{\pi}$} & $\norm{\mbf{r}}_F$             & 5.4222e-04 & 2.8233e-05 & 5.1908e-07 & 1.1113e-08 & 7.3918e-11 \\                                             
 &                                $\norm{\msf{Q} - \msf{I}_D}_F$ & 5.2358e-04 & 2.0300e-05 & 4.4962e-07 & 7.0963e-09 & 5.1610e-11 \\ \cline{2-7}                                 
 &                                $N$ & \multicolumn{1}{c}{11} & \multicolumn{1}{c}{12} & \multicolumn{1}{c}{13} & \multicolumn{1}{c}{14} & \multicolumn{1}{c}{15} \\ \cline{2-7}
 &                                $\norm{\mbf{r}}_F$             & 2.7305e-13 & 1.2477e-14 & 6.2789e-15 & 5.8445e-15 & 7.2210e-15 \\                                             
 &                                $\norm{\msf{Q} - \msf{I}_D}_F$ & 1.6628e-13 & 4.3047e-16 & 4.1495e-16 & 2.8105e-16 & 2.2323e-16 \\ \hline                                      
$\sigma$ &                        $N$ & \multicolumn{1}{c}{6} & \multicolumn{1}{c}{7} & \multicolumn{1}{c}{8} & \multicolumn{1}{c}{9} & \multicolumn{1}{c}{10} \\ \hline         
\multirow{5}{*}{$4 \sqrt{\pi}$} & $\norm{\mbf{r}}_F$             & 4.5343e-01 & 2.6201e-01 & 7.5299e-02 & 3.4864e-02 & 6.9207e-03 \\                                             
 &                                $\norm{\msf{Q} - \msf{I}_D}_F$ & 4.3784e-01 & 1.8839e-01 & 6.5223e-02 & 2.2263e-02 & 4.8319e-03 \\ \cline{2-7}                                 
 &                                $N$ & \multicolumn{1}{c}{11} & \multicolumn{1}{c}{12} & \multicolumn{1}{c}{13} & \multicolumn{1}{c}{14} & \multicolumn{1}{c}{15} \\ \cline{2-7}
 &                                $\norm{\mbf{r}}_F$             & 1.0987e-03 & 7.8620e-05 & 1.0067e-14 & 1.2076e-14 & 1.2964e-14 \\                                             
 &                                $\norm{\msf{Q} - \msf{I}_D}_F$ & 6.6716e-04 & 6.9627e-05 & 6.0970e-16 & 7.7092e-16 & 1.0855e-15 \\                                             

\hline
\end{tabular}
\caption{\label{tabD} Comparison of the norm of the residual of the Gabor transform with width $\tau = 12$ and decay $\sigma$ indicated to the distance from the quality matrix $\msf{Q}$ to the identity $\msf{I}_D$ at various orders $N$ and even parity for an irregularly sampled signal with $D = 30$ and $T = 100$ \revise{such that $N_D = 15$ and $N_T = 50$.}}
\end{table}

For the Morlet basis functions, the construction of $\msf{\Psi}$ is more involved, as $\Psi(t',n) = \Phi(t',n) \Theta(t',n)$.  Nonetheless, one can evaluate $\msf{Q}$ from $\msf{\Psi}^\ast \msf{D}_f \msf{\Psi}^\dagr$ for any rank $T$.  As a function of order $N$, one finds that the deviation $\norm{\msf{Q} - \msf{I}_T}_F$ decreases until $N > T/2$, after which it bottoms out on the order of a few percent times $T$.  In Table~\ref{tabE} we compare the norm of the residual for the Morlet basis using $\sigma = \sqrt{\pi}$ and $\tau = 6$ for some regularly sampled signal with $D = T = 30$ to the deviation of $\msf{Q}$ from $\msf{I}_D$ at various $N$ with $P = 0$.  If one wishes to investigate the feasibility of devising a Morlet basis with perfect reconstruction, the evaluation of $\msf{Q}$ is where to start.

\begin{table}
\centering
\begin{tabular}{c|cccccccccc}
\hline
$N$ & \multicolumn{1}{c}{6} & \multicolumn{1}{c}{7} & \multicolumn{1}{c}{8} & \multicolumn{1}{c}{9} & \multicolumn{1}{c}{10} \\ \hline    
$\norm{\mbf{r}}_F$             & 3.3471e+00 & 2.9397e+00 & 2.7428e+00 & 2.2582e+00 & 1.9475e+00 \\                                        
$\norm{\msf{Q} - \msf{I}_D}_F$ & 2.9527e+00 & 2.4686e+00 & 2.0993e+00 & 1.8244e+00 & 1.5788e+00 \\ \hline                                 
$N$ & \multicolumn{1}{c}{11} & \multicolumn{1}{c}{12} & \multicolumn{1}{c}{13} & \multicolumn{1}{c}{14} & \multicolumn{1}{c}{15} \\ \hline
$\norm{\mbf{r}}_F$             & 1.7911e+00 & 1.5917e+00 & 1.3310e+00 & 9.3137e-01 & 5.3677e-01 \\                                        
$\norm{\msf{Q} - \msf{I}_D}_F$ & 1.3451e+00 & 1.1139e+00 & 8.7648e-01 & 6.3175e-01 & 4.0199e-01 \\                                        

\hline
\end{tabular}
\caption{\label{tabE} Comparison of the norm of the residual of the Morlet transform \revise{with width $\tau = 6$ and decay $\sigma = \sqrt{\pi}$} to the distance from the quality matrix $\msf{Q}$ to the identity $\msf{I}_D$ at various orders $N$ and even parity for a regularly sampled signal with $D = T = 30$ \revise{such that $N_D = N_T = 15$.}}
\end{table}

\begin{figure}
\centering
\includegraphics[]{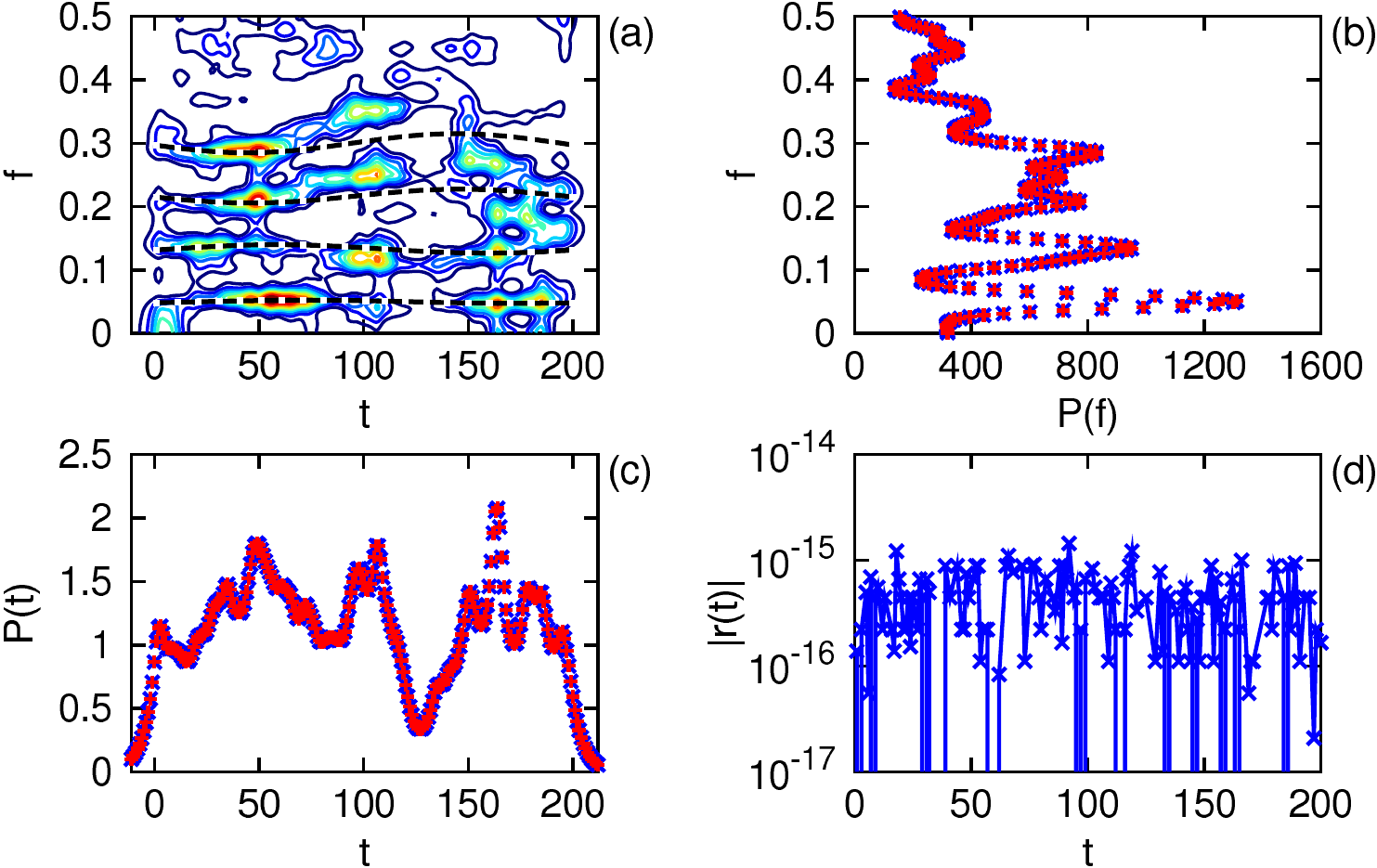}%
\caption{\label{figI} \revise{One-sided layered window transform of an irregularly sampled signal as described in the text.  The power spectrum is shown in (a) with the signal's instantaneous frequencies indicated by the dashed lines.  The marginal spectral power is shown in (b) as $\times$, as is the marginal temporal power in (c), and each is compared to the convolution of the window and signal energies indicated by $+$.  The absolute value of the reconstruction residual is displayed in (d).}}
\end{figure}

Let us close this section by looking at the analysis of the signal from the previous section but with the number of measurements reduced by a factor of a third ($N_D = 134$) for the same duration $T = 200$.  The evaluation of the layered window Fourier transform proceeds as before, with the understanding that values of $y(\hat{t} + t')$ at times not in $\mbf{t}$ are treated as zero; the time axis for the spectrum $\hat{Y}(n, \hat{t})$ includes every integer $\hat{t} \in [1-\tau_L, T+\tau_L]$.  Using the same parameters as before, $\sigma = \sqrt{\pi} / 6$ and $\tau \in [12, 96]$ with $N = 200$ and $P = 0$, in Figure~\ref{figI} we display the power spectrum, its marginal densities, and the reconstruction residual for our irregularly sampled signal.  While the loss of information has affected the appearance of the spectrum relative to the regularly sampled case, it nonetheless remains a faithful representation of the information which is available.  By virtue of using a single fixed window $\Phi_L$, the layered window Fourier transform satisfies the fundamental theorems of spectral analysis even when there are gaps in the observation record.

%%%%%%%%%%%%%%%%%%%%%%%%%%%%%%%%%%%%%%%%%%%%%%%%%%%%%%%%%%%%

\section{Window Comparison}

Let us conclude the analysis with a comparison of the temporal and spectral bandwidths for several types of window function, as well as their estimates of the power spectral density carried by a test signal with a little more complexity than we had before.  The test signal is regularly sampled with $T = 200$ and four component frequencies, but now only two components are present in the first half of the duration, while the other two are in the second half.  An abrupt transition occurs at the midpoint of the duration.  The amplitude of the frequency variation is now 10\%, with a period of $T/2$, so that each component covers a broad range of instantaneous frequency.  The order for all the transforms in this section will be the minimal Fourier order of the signal $N = 100$ with even parity $P = 0$.

\begin{figure}
\centering
\includegraphics[]{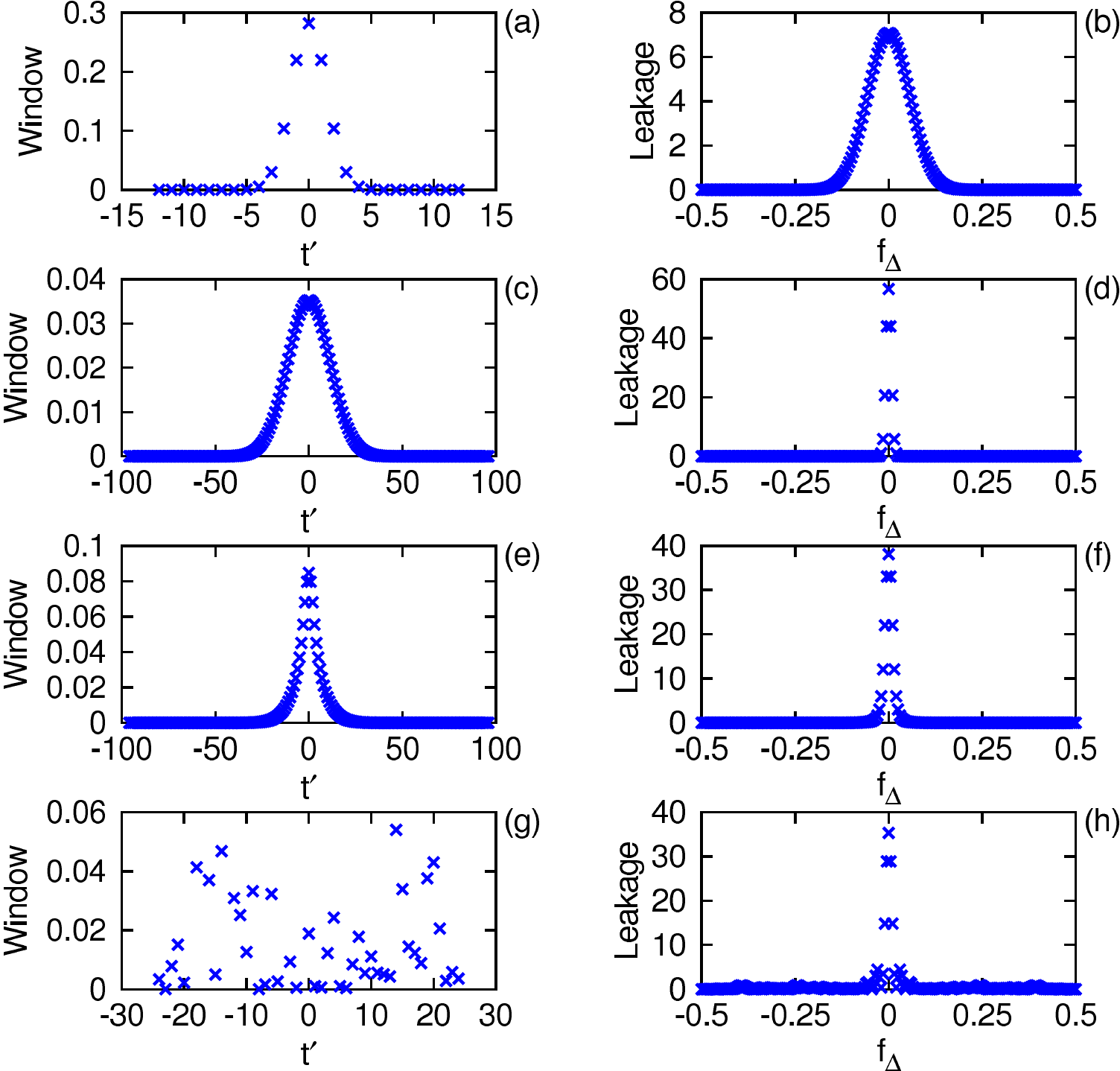}%
\caption{\label{figK} Comparison of the temporal and spectral energy densities of the various windows arranged according to Table~\ref{tabF}.}
\end{figure}

\begin{table}
\centering
\begin{tabular}{c|ccrrrrr|c}
\hline
Window & $\sigma$ & $\tau$ & \multicolumn{1}{c}{$t_1$} & \multicolumn{1}{c}{$t_2$} & \multicolumn{1}{c}{$f_1$} & \multicolumn{1}{c}{$f_2$} & \multicolumn{1}{c|}{$4 \pi t_2 f_2$} & $\norm{\mbf{r}}_F$ \\ 
\hline
Gabor & $2 \sqrt{\pi}$ & $12$ & -0.000 & 1.414 & -0.000 & 0.056 & 1.000 & 0.000 \\
Gabor & $16 \sqrt{\pi}$ & $96$ & -0.000 & 11.314 & 0.000 & 0.007 & 1.000 & 0.000 \\
Layered & $\sqrt{\pi} / 6$ & $[12, 96]$ & 0.000 & 6.990 & 0.000 & 0.014 & 1.244 & 0.000 \\
Random & --- & $24$ & -2.836 & 13.718 & 0.000 & 0.130 & 22.442 & 0.000 \\

\hline
\end{tabular}
\caption{\label{tabF} Comparison of the temporal and spectral bandwidths for various windows, as well as their rms reconstruction residual for the test signal as described in the text.}
\end{table}

The parameters for the four types of window considered are displayed in Table~\ref{tabF}, comprising of two Gabor windows at either end of the range in scale for the layered window also considered, in addition to a random window with an arbitrary duration.  The first moments of time and frequency are indicated by $t_1$ and $f_1$ respectively, and the bandwidths (square root of the second moment about the first moment) by $t_2$ and $f_2$.  The two Gabor windows are observed to minimize the Fourier uncertainty relation $t_2 f_2 \geq 1 / 4 \pi$, thus in that sense are optimal, while the layered window has a slightly larger bandwidth product.  The random window has a bandwidth product which is considerably larger, yet all these windows produce a valid discrete transform pair that satisfies the fundamental theorems of spectral analysis as indicated by the residual of reconstruction $\norm{\mbf{r}}_F$.  The energy densities in the temporal and spectral representation used to evaluate the bandwidth products are shown in Figure~\ref{figK}, where one can see the trade-off in time/frequency resolution in action.

\begin{figure}
\centering
\includegraphics[]{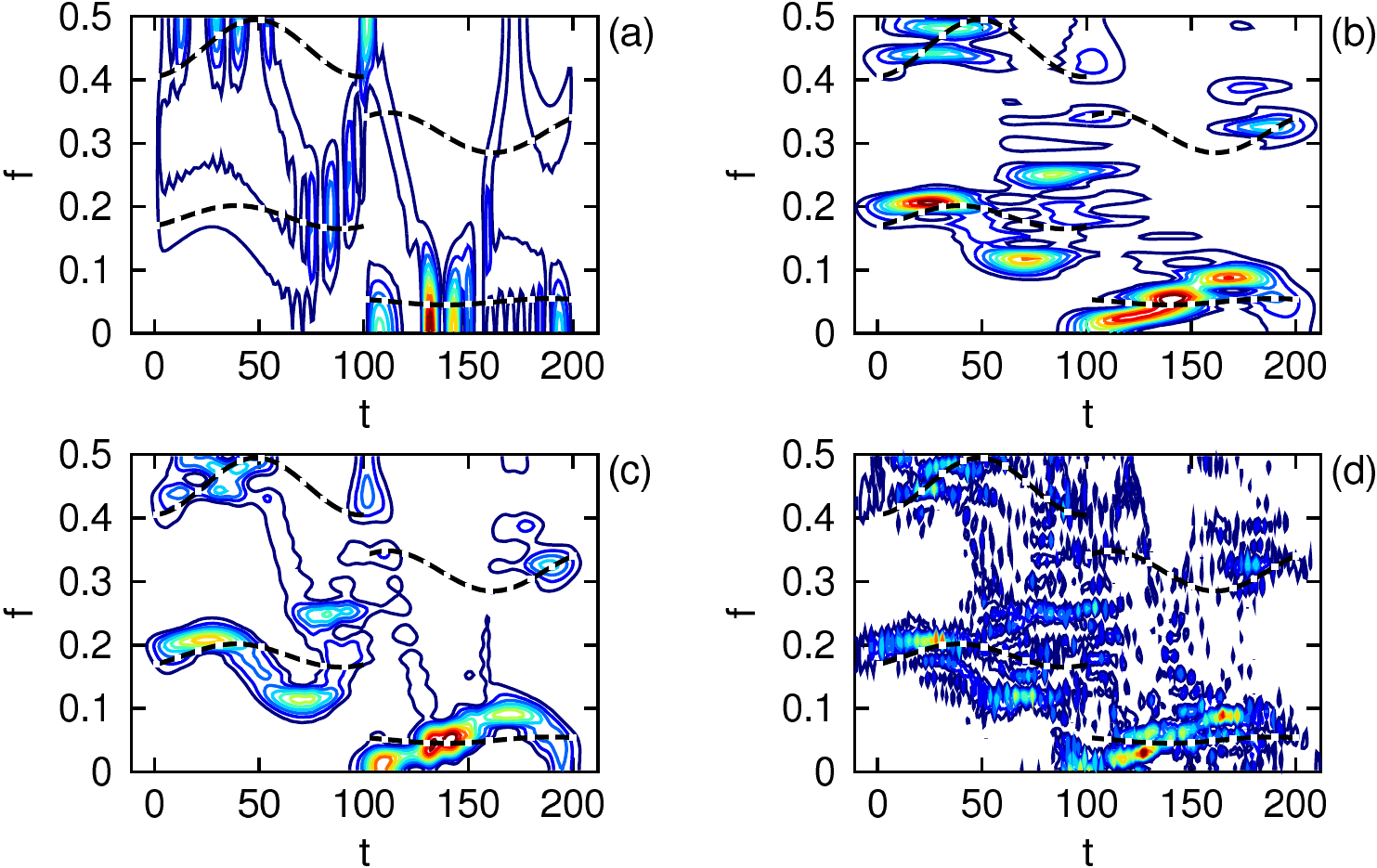}%
\caption{\label{figL} Comparison of the power spectral density of the test signal described in the text using the various windows arranged according to Table~\ref{tabF}.  The instantaneous frequencies used to generate the signal are indicated by the dashed lines.}
\end{figure}

The power spectral density estimates evaluated from the test signal of this section using the various windows are displayed in Figure~\ref{figL}, arranged from (a) to (d) according to Table~\ref{tabF}.  The power spectrum given by the layered window in (c) does indeed incorporate features of either Gabor transform shown in (a) and (b); in a sense, it has combined the Gabor transforms over its range of scale in a way that preserves the energy and reconstruction theorems.  With some algebraic manipulation of the expressions defining the layered window Fourier transform one might be able to show that explicitly, but for now that statement is intuitive speculation.  Interestingly, the spectrum given by the random window is not much different from the others, driving home the point that the shape of the window truly is arbitrary as long as it is normalized appropriately.

%%%%%%%%%%%%%%%%%%%%%%%%%%%%%%%%%%%%%%%%%%%%%%%%%%%%%%%%%%%%

\section{Discussion}

The primary result of this investigation is that the windowed Fourier transform has far more flexibility and utility than it is usually credited with.  To achieve satisfaction of the energy and reconstruction theorems in the discrete setting, \revise{the only requirements on the window are that it be real, nonnegative, and normalized explicitly to unit energy, or two units for a one-sided transform.}  While we have looked only at symmetric windows here, one can easily verify that the expressions for the transform pair, either Equations~(\ref{eqnB}-\ref{eqnC}) or (\ref{eqnD}-\ref{eqnE}), remain valid for a window that is not symmetric in $t'$; with attention to the details of relocating the temporal bins, the coefficients can be assigned to the time corresponding to the peak of the window rather than its midpoint.  Similarly, a window with an even length duration $T'$ could be used if one is careful with the definition of the phase and the location of the bins.  In fact, any nonnegative function can be used for $\Phi$, or even a list of arbitrary numbers, as long as it is suitably normalized.

The flexibility in $\Phi$ allows one to define a multiresolution spectral analysis in terms of an atomic unit of energy $\Phi_A(t/\tau)$ evaluated over a range of scales $\tau$.  These multiple scalings of $\Phi_A$ are combined into a single layered window $\Phi_L$ which is applied to the entire frequency axis; the phase component of the basis is independent of the window function.  Because only a single window is used, the fundamental theorems of spectral analysis are satisfied.  With a Gaussian $\Phi_A$, the form of $\Phi_L$ very closely resembles a Lorentzian function, which expresses uniformity over scale when expressed as an angle.  The shape of $\Phi_L$ can be tailored to the needs of the investigation through inspection of its leakage function.  Similarly to the wavelet transform, the trade-off between resolution in time or frequency is under the control of the investigator.  \revise{Any structure seen in the spectral coefficients of the signal is understood to be conditioned on the selection of the window function;} the answer one gets depends upon how the question is asked.

The difficulties faced by the continuous wavelet transform exemplified by the Morlet basis, with respect to the fundamental theorems of spectral analysis, can be summarized in the evaluation of its quality matrix $\msf{Q}$.  Perfect reconstruction in the discrete setting requires $\msf{Q}$ to equal the identity matrix to the precision of one's computational device.  The various adjustments looked at above which improve the reconstruction residual must be bringing $\msf{Q}$ closer to that form; however, none of them achieve the required precision.  For an integral transform pair to have quantitative significance, one must demonstrate that the distance from its $\msf{Q}$ to $\msf{I}$ is numerically zero.  This investigation has proposed an alternate approach to multiresolution analysis which does satisfy the energy and reconstruction theorems while improving upon the resolution properties of the Gabor transform.

The effects of aliasing and of irregular sampling are easily understood with regard to the periodicity induced in the Fourier spectrum.  The Nyquist interval is given simply by the span between frequencies which are indistinguishable over the stated measurement times; assigning a value of unity to that range such that $f_c = 1/2$ is just a matter of choosing the appropriate units for time.  The implementation of the windowed Fourier transform is unaffected beyond zero padding the signal at locations of missing measurements, whose justification is that no other procedure leaves the energy, hence information content, unchanged.  One should observe that the ends of a regularly sampled signal are treated the same way.  On that note, we can remark that there is no cone of influence for the windowed Fourier transform, as every spectral coefficient is important to the satisfaction of the fundamental theorems, \revise{even those outside the temporal domain of the data.}

The minimal order $N_\mathrm{min}$ required for a faithful representation of the signal depends upon the temporal bandwidth of the window and the duration of the data.  Pinning that number down for the windowed Fourier transform is a bit tricky, and if needed is best evaluated explicitly through inspection of $\msf{Q}$ as a function of order $N$.  What one can say in general is that $1 \leq N_\mathrm{min} \leq N_T$, where the bounds are given by the minimal orders for the fully temporal and fully spectral representations of the regularly sampled signal, respectively.  Irregular sampling complicates matters by introducing an order $N_D < N_T$ such that $N_D \leq N_\mathrm{min} \leq N_T$ \revise{for the full-length Fourier transform and $N_\mathrm{min} \leq N_D$ for the windowed Fourier transform}.  For most practical cases of data analysis, one is interested in producing a smooth plot of the spectral content, thus one would use an order $N \gg N_\mathrm{min}$.  To verify the marginal density of the windowed power spectrum in comparison to the convolution of the window and signal energy densities in the spectral representation, one requires an order $N$ sufficient to resolve both the signal and the window temporal durations.

\begin{figure}
\centering
\includegraphics[]{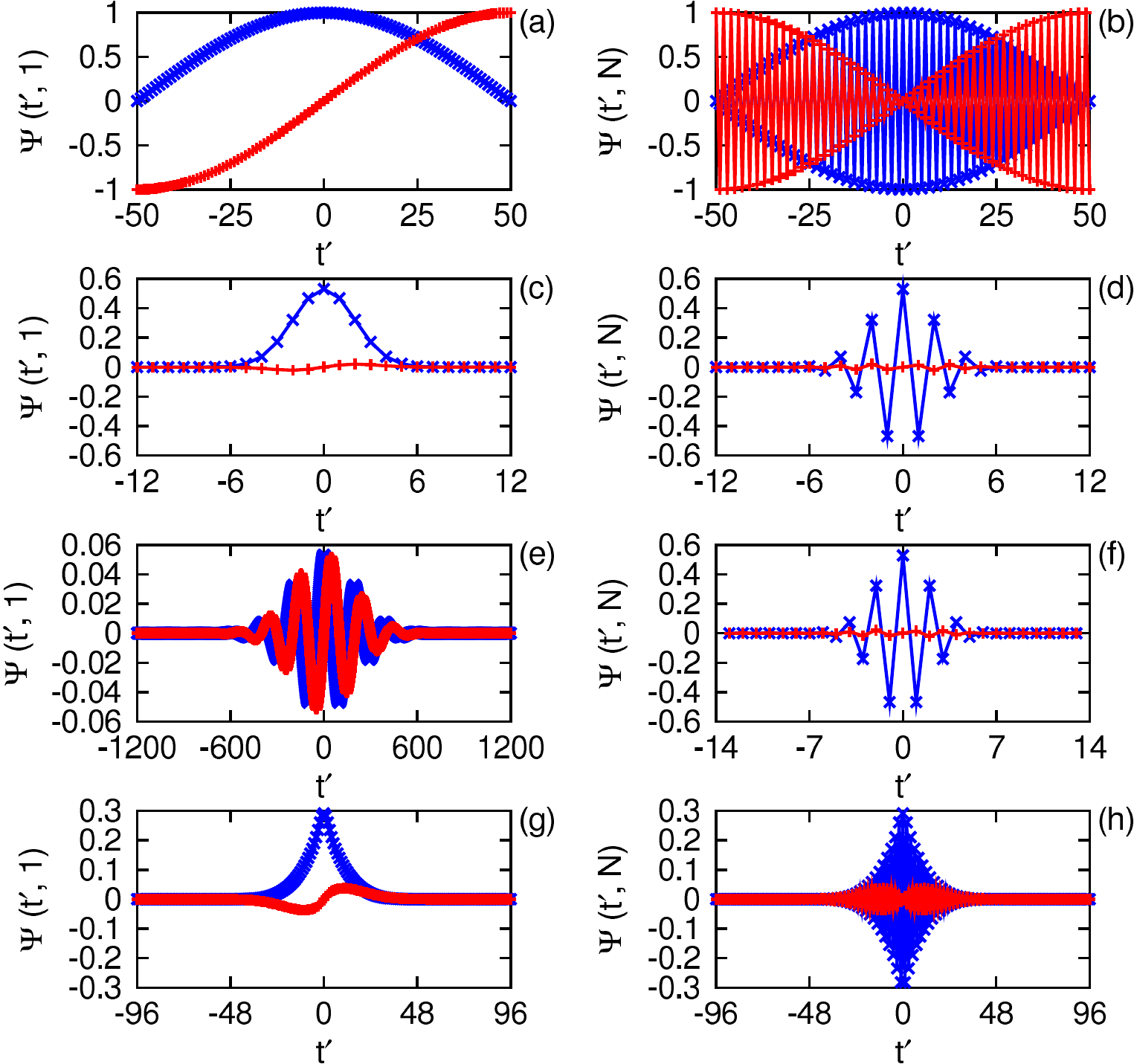}%
\caption{\label{figJ} \revise{Comparison of the lowest $\Psi(t',1)$ and highest $\Psi(t',N)$ frequency basis functions at order $N = 50$ with $P = 0$ using the parameters found in the text, with the real part indicated by $\times$ and the imaginary part by $+$.  The Fourier basis is in (a) and (b), the Gabor basis is in (c) and (d), the Morlet basis is in (e) and (f), and the layered window basis is in (g) and (h).}}
\end{figure}

Thus far we have heard nary a peep from the actual basis functions used in this analysis, so let us close the discussion by looking at some of the stars of the show.  In Figure~\ref{figJ} we display \revise{the real and imaginary parts of} the basis functions, normalized to unit energy, for each of the transforms considered above at order $N = 50$ with even parity $P = 0$ at the lowest and highest entries on the frequency axis, \textit{i.e.} $f_1 = 1 / 200$ and $f_N = 99 / 200$.  The Fourier transform has no intrinsic window, so it is shown for a duration $T' = 101$ which spans $T < T'$.  The window parameters chosen are those from the previous sections: $\sigma = 2 \sqrt{\pi}$ and $\tau = 12$ for the Gabor transform, $\sigma = \sqrt{\pi}$ and $\tau = 6$ for the Morlet transform, and $\sigma = \sqrt{\pi} / 6$ and $\tau \in [12, 96]$ for the layered window transform.  One can observe that the transforms whose $\msf{Q}$ equals the identity share the property that these basis functions are phase analogues over the discrete sample times, whereas the Morlet basis functions are scale analogues.  By phase analogue we mean if $\Psi(t',1) = a_1 + i b_1$, then $\Psi(t',N) = a_1 (-1)^{t'} + i b_1 (-1)^{t'+1}$, a feature shared by all positive frequency pairs whose midpoint is $f = 1/4$; the Morlet basis differs from the others in this regard.  Whether that property is required for satisfaction of the fundamental theorems is unclear, but the results of this analysis suggest that it is.

%%%%%%%%%%%%%%%%%%%%%%%%%%%%%%%%%%%%%%%%%%%%%%%%%%%%%%%%%%%%

\section{Conclusions}

In this article we have compared the Fourier, Gabor, and Morlet transforms in the fully discrete setting applicable to the analysis of sampled data.  While the Fourier and Gabor transforms satisfy the fundamental theorems of energy conservation and perfect reconstruction, the Morlet transform does not.  The magnitude of the residual can be related to the distance from the quality matrix to the identity in all cases, such that the minimal order for reconstruction can be determined for the Fourier and Gabor transforms.  Various methods of improving the response of the Morlet transform are considered; however, none of them achieve the desired precision on par with the truncation error of one's computational device.

An alternate approach to multiresolution analysis is proposed which constructs a single layered window from multiple scalings of some atomic unit of energy.  This layered window transform satisfies the fundamental theorems of spectral analysis while providing sufficient temporal resolution to identify non-stationary features in the signal.  The trade-off between time and frequency resolution is under the control of the investigator through the selection of the atomic window and the scales over which it is evaluated.  The power spectrum of the layered window transform is similar to that of the Morlet transform but provides much better frequency resolution.

The premise behind the wavelet transform is that the low frequency elements of a signal should have a much longer duration than the high frequency elements.  There are, however, many examples of real world signals for which the converse is true.  Consider, for example, the digital recording of a kick drum and cymbal rhythm such that the low frequency bursts have a relatively short duration compared to the ringing at high frequency.  For that type of signal, the wavelet transform is going to provide a poor choice of basis even if it satisfied the fundamental theorems.  The flexibility in the choice of window function used in the windowed Fourier transform allows one to tailor its response to the needs of the analysis far better.  The result one gets depends upon how one defines the resolution of the window, any of which provide a valid spectral representation of the signal.

To make quantitative use of the spectral analysis of some discretely sampled signal, one must demonstrate that the fundamental theorems are satisfied.  The windowed Fourier transform can be shown to satisfy those requirements for any real valued window function that is suitably normalized.  The potential for gaps in the data, or some other form of irregular sampling, is found to pose no problem once the correct Nyquist interval is identified.  Consequently, the windowed Fourier transform can be applied to data from a wide variety of sources, such as astronomical observations, which are limited physically to an irregular set of observation times, as well as the common case of regular sampling.  The implementation of a multiresolution spectral analysis in the discrete setting appears to be possible only when the multiple scalings of the energy distribution are applied evenly across the frequency axis by combining them into a single layered window.

%%%%%%%%%%%%%%%%%%%%%%%%%%%%%%%%%%%%%%%%%%%%%%%%%%%%%%%%%%%%

%\section*{Acknowledgements}
%
%Main text.

%==========================================================
%==========================================================
% Back Matter (References and Notes)
%----------------------------------------------------------
% Style and layout of the references
\bibliographystyle{mdpi}
\makeatletter
\renewcommand\@biblabel[1]{#1. }
\makeatother
%----------------------------------------------------------
%Citing a journal paper \cite{ref-journal}. And now citing a book reference \cite{ref-book}.
% Use the following option to include external BibTeX files:
%\bibliography{../wavelet,../solar,../stats}

\begin{thebibliography}{-------}
\providecommand{\natexlab}[1]{#1}

\bibitem[Torrence and Compo(1998)]{torrence:98}
Torrence, C.; Compo, G.P.
\newblock A Practical Guide to Wavelet Analysis.
\newblock {\em Bulletin of the American Meteorological Society} {\bf 1998},
  {\em 79},~61--78.

\bibitem[Farge(1992)]{farge-1992}
Farge, M.
\newblock Wavelet Transforms and their Applications to Turbulence.
\newblock {\em Annual Review of Fluid Mechanics} {\bf 1992}, {\em
  24},~395--458.

\bibitem[Grossmann and Morlet(1984)]{grossman-1984}
Grossmann, A.; Morlet, J.
\newblock Decomposition of Hardy Functions into Square Integrable Wavelets of
  Constant Shape.
\newblock {\em SIAM Journal on Mathematical Analysis} {\bf 1984}, {\em
  15},~723--736.

\bibitem[Meyer(1986-1987)]{Meyer-1986}
Meyer, Y.
\newblock Ondelettes et Fonctions Splines.
\newblock {\em Séminaire Équations aux dérivées partielles (dit
  "Goulaouic-Schwartz")} {\bf 1986-1987}, pp. 1--18.

\bibitem[Mallat(1989)]{mallat-192463}
Mallat, S.
\newblock A Theory for Multiresolution Signal Secomposition: {T}he Wavelet
  Representation.
\newblock {\em Pattern Analysis and Machine Intelligence, IEEE Transactions on}
  {\bf 1989}, {\em 11},~674 --693.

\bibitem[Daubechies(1988)]{daubechies-1988}
Daubechies, I.
\newblock Orthonormal Bases of Compactly Supported Wavelets.
\newblock {\em Communications on Pure and Applied Mathematics} {\bf 1988}, {\em
  41},~909--996.

\bibitem[Vetterli and Herley(1992)]{vetterli-157221}
Vetterli, M.; Herley, C.
\newblock Wavelets and Filter Banks: {T}heory and Design.
\newblock {\em Signal Processing, IEEE Transactions on} {\bf 1992}, {\em
  40},~2207 --2232.

\bibitem[Kronland-Martinet \em{et~al.}(1987)Kronland-Martinet, Morlet, and
  Grossmann]{k-mmg-1987gs}
Kronland-Martinet, R.; Morlet, J.; Grossmann, A.
\newblock Analysis Of Sound Patterns Through Wavelet Transforms.
\newblock {\em International Journal of Pattern Recognition and Artificial
  Intelligence} {\bf 1987}, {\em 01},~273--302.

\bibitem[{Meyers} \em{et~al.}(1993){Meyers}, {Kelly}, and
  {O'Brien}]{meyers-1993}
{Meyers}, S.D.; {Kelly}, B.G.; {O'Brien}, J.J.
\newblock {An Introduction to Wavelet Analysis in Oceanography and Meteorology:
  With Application to the Dispersion of Yanai Waves}.
\newblock {\em Monthly Weather Review} {\bf 1993}, {\em 121},~2858.

\bibitem[{Baliunas} \em{et~al.}(1997){Baliunas}, {Frick}, {Sokoloff}, and
  {Soon}]{Baliunas:grl2411}
{Baliunas}, S.; {Frick}, P.; {Sokoloff}, D.; {Soon}, W.
\newblock Time Scales and Trends in the {Central England Temperature} Data
  (1659-1990): A Wavelet Analysis.
\newblock {\em Geophys. Res. Lett.} {\bf 1997}, {\em 24},~1351--1354.

\bibitem[{Fligge} \em{et~al.}(1999){Fligge}, {Solanki}, and {Beer}]{fligge-313}
{Fligge}, M.; {Solanki}, S.K.; {Beer}, J.
\newblock Determination of Solar Cycle Length Variations Using the Continuous
  Wavelet Transform.
\newblock {\em Astron. Astrophys.} {\bf 1999}, {\em 346},~313--321.

\bibitem[Christopoulou \em{et~al.}(2002)Christopoulou, Skodras, and
  Georgakilas]{ans-c49}
Christopoulou, E.B.; Skodras, A.N.; Georgakilas, A.A.
\newblock Time series analysis of sunspot oscillations using the wavelet
  transform.
\newblock {\em Digital Signal Processing, 2002. DSP 2002. 2002 14th
  International Conference on} {\bf 2002}, {\em 2},~893--896.

\bibitem[Le and Wang(2003)]{cjaa:35391}
Le, G.M.; Wang, J.L.
\newblock Wavelet Analysis of Several Important Periodic Properties in the
  Relative Sunspot Numbers.
\newblock {\em Chin. J. Astron. Astrophys.} {\bf 2003}, {\em 3},~391--394.

\bibitem[Grinsted \em{et~al.}(2004)Grinsted, Moore, and
  Jevrejeva]{npg-11-561-2004}
Grinsted, A.; Moore, J.C.; Jevrejeva, S.
\newblock Application of the cross wavelet transform and wavelet coherence to
  geophysical time series.
\newblock {\em Nonlinear Processes in Geophysics} {\bf 2004}, {\em
  11},~561--566.

\bibitem[Piscaronoft \em{et~al.}(2004)Piscaronoft, Kalvov\'{a}, and
  Br\'{a}zdil]{piscaron-1661}
Piscaronoft, P.; Kalvov\'{a}, J.; Br\'{a}zdil, R.
\newblock Cycles and trends in the {C}zech temperature series using wavelet
  transforms.
\newblock {\em International Journal of Climatology} {\bf 2004}, {\em
  24},~1661--1670.

\bibitem[Liu \em{et~al.}(2007)Liu, Liang, and Weisberg]{liu-2093}
Liu, Y.; Liang, X.S.; Weisberg, R.H.
\newblock Rectification of the Bias in the Wavelet Power Spectrum.
\newblock {\em J. Atmos. Oceanic Technol.} {\bf 2007}, {\em 24},~2093--2102.

\bibitem[Jevrejeva \em{et~al.}(2003)Jevrejeva, Moore, and Grinsted]{jgr-a-2156}
Jevrejeva, S.; Moore, J.C.; Grinsted, A.
\newblock Influence of the Arctic Oscillation and El Niño-Southern Oscillation
  (ENSO) on ice conditions in the Baltic Sea: The wavelet approach.
\newblock {\em Journal of Geophysical Research: Atmospheres} {\bf 2003}, {\em
  108},~4677.

\bibitem[Echer \em{et~al.}(2009)Echer, Echer, Nordemann, and Rigozo]{echer-41}
Echer, M.P.S.; Echer, E.; Nordemann, D.J.R.; Rigozo, N.R.
\newblock Multi-resolution analysis of global surface air temperature and solar
  activity relationship.
\newblock {\em Journal of Atmospheric and Solar-Terrestrial Physics} {\bf
  2009}, {\em 71},~41--44.

\bibitem[Greene(2008)]{greene-2273}
Greene, N.
\newblock Inverse Wavelet Reconstruction for Resolving the {G}ibbs Phenomenon.
\newblock {\em International Journal of Circuits, Systems and Signal
  Processing} {\bf 2008}, {\em 2},~73--77.

\bibitem[Press \em{et~al.}(1992)Press, Teukolsky, Vetterling, and
  Flannery]{Press-1992}
Press, W.; Teukolsky, S.; Vetterling, W.; Flannery, B.
\newblock {\em Numerical Recipes in C}, 2nd ed.; Cambridge University Press:
  Cambridge, England,  1992.

\bibitem[Sadowsky(1996)]{Sadowsky:1996}
Sadowsky, J.
\newblock Investigation of Signal Characteristics Using the Continuous Wavelet
  Transform.
\newblock {\em Johns Hopkins APL Technical Digest} {\bf 1996}, {\em
  17},~258--269.

\bibitem[Johnson(2012)]{rwj:nova01}
Johnson, R.W., Wavelets: Classification, Theory and Applications; Nova Science
  Publishers: Hauppauge, NY,  2012; Chapter 6. Extended Wavelet
  Transform for Discretely Sampled Data, pp. 125--155.

\bibitem[{Johnson}(2012)]{rwj:ijwmip01}
{Johnson}, R.W.
\newblock Symmetrization and enhancement of the continuous {Morlet} transform
  for spectral density estimation.
\newblock {\em International Journal of Wavelets, Multiresolution and
  Information Processing} {\bf 2012}, {\em 10},~1250009.

\bibitem[Lomb(1976)]{Lomb-447}
Lomb, N.R.
\newblock Least-squares frequency analysis of unequally spaced data.
\newblock {\em Astrophysics and Space Science} {\bf 1976}, {\em 39},~447--462.

\bibitem[{Scargle}(1982)]{Scargle-835}
{Scargle}, J.D.
\newblock {Studies in astronomical time series analysis. II - {S}tatistical
  aspects of spectral analysis of unevenly spaced data}.
\newblock {\em The Astrophysical Journal} {\bf 1982}, {\em 263},~835--853.

\bibitem[Foster(1996)]{Foster:1996}
Foster, G.
\newblock Wavelets For Period Analysis of Unevenly Samples Time Series.
\newblock {\em {The} Astronomical Journal} {\bf 1996}, {\em 112},~1709--1729.

\bibitem[Frick \em{et~al.}(1997)Frick, Baliunas, Galyagin, Sokoloff, and
  Soon]{Frick:1997426}
Frick, P.; Baliunas, S.L.; Galyagin, D.; Sokoloff, D.; Soon, W.
\newblock Wavelet Analysis of Stellar Chromospheric Activity Variations.
\newblock {\em The Astrophysical Journal} {\bf 1997}, {\em 483},~426--434.

\bibitem[{Johnson}(2010)]{rwj:astro01}
{Johnson}, R.W.
\newblock Edge adapted wavelets, solar magnetic activity, and climate change.
\newblock {\em Astrophysics and Space Science} {\bf 2010}, {\em 326},~181--189.

\bibitem[Sweldens(1998)]{sweldens98lifting}
Sweldens, W.
\newblock The Lifting Scheme: {A} Construction of Second Generation Wavelets.
\newblock {\em SIAM Journal on Mathematical Analysis} {\bf 1998}, {\em
  29},~511--546.

\bibitem[Johnson(2012)]{rwj:astro03}
Johnson, R.W.
\newblock Max{E}nt power spectrum estimation using the {Fourier} transform for
  irregularly sampled data applied to a record of stellar luminosity.
\newblock {\em Astrophysics and Space Science} {\bf 2012}, {\em 338},~35--48.

\end{thebibliography}
%----------------------------------------------------------

\end{document}